\documentstyle[epsfig]{mn}
\newif\ifAMStwofonts

\ifoldfss
  \ifCUPmtlplainloaded \else
    \NewTextAlphabet{textbfit} {cmbxti10} {}
    \NewTextAlphabet{textbfss} {cmssbx10} {}
    \NewMathAlphabet{mathbfit} {cmbxti10} {} 
    \NewMathAlphabet{mathbfss} {cmssbx10} {} 
  \fi
  \ifAMStwofonts
    \ifCUPmtlplainloaded \else
      \NewSymbolFont{upmath} {eurm10}
      \NewSymbolFont{AMSa} {msam10}
      \NewMathSymbol{\upi}     {0}{upmath}{19}
      \NewMathSymbol{\umu}     {0}{upmath}{16}
      \NewMathSymbol{\upartial}{0}{upmath}{40}
      \NewMathSymbol{\leqslant}{3}{AMSa}{36}
      \NewMathSymbol{\geqslant}{3}{AMSa}{3E}

      \let\geq=\geqslant 
    \fi
  \fi
\fi 

\ifnfssone
  \newmathalphabet{\mathit}
  \addtoversion{normal}{\mathit}{cmr}{m}{it}
  \addtoversion{bold}{\mathit}{cmr}{bx}{it}
  \newmathalphabet{\mathbfit} 
  \addtoversion{normal}{\mathbfit}{cmr}{bx}{it}
  \addtoversion{bold}{\mathbfit}{cmr}{bx}{it}
  \newmathalphabet{\mathbfss} 
  \addtoversion{normal}{\mathbfss}{cmss}{bx}{n}
  \addtoversion{bold}{\mathbfss}{cmss}{bx}{n}
  \ifAMStwofonts
    \ifCUPmtlplainloaded \else
      \UseAMStwoboldmath
      \makeatletter
      \new@mathgroup\upmath@group
      \define@mathgroup\mv@normal\upmath@group{eur}{m}{n}
      \define@mathgroup\mv@bold\upmath@group{eur}{b}{n}
      \edef\UPM{\hexnumber\upmath@group}
      \new@mathgroup\amsa@group
      \define@mathgroup\mv@normal\amsa@group{msa}{m}{n}
      \define@mathgroup\mv@bold\amsa@group{msa}{m}{n}
      \edef\AMSa{\hexnumber\amsa@group}
      \makeatother
      \mathchardef\upi="0\UPM19
      \mathchardef\umu="0\UPM16
      \mathchardef\upartial="0\UPM40
      \mathchardef\leqslant="3\AMSa36
      \mathchardef\geqslant="3\AMSa3E

      \let\geq=\geqslant 
    \fi
  \fi
\fi 

\ifnfsstwo
  \DeclareMathAlphabet{\mathbfit}{OT1}{cmr}{bx}{it}
  \SetMathAlphabet\mathbfit{bold}{OT1}{cmr}{bx}{it}
  \DeclareMathAlphabet{\mathbfss}{OT1}{cmss}{bx}{n}
  \SetMathAlphabet\mathbfss{bold}{OT1}{cmss}{bx}{n}
  \ifAMStwofonts
    \ifCUPmtlplainloaded \else
      \DeclareSymbolFont{UPM}{U}{eur}{m}{n}
      \SetSymbolFont{UPM}{bold}{U}{eur}{b}{n}
      \DeclareSymbolFont{AMSa}{U}{msa}{m}{n}
      \DeclareMathSymbol{\upi}{0}{UPM}{"19}
      \DeclareMathSymbol{\umu}{0}{UPM}{"16}
      \DeclareMathSymbol{\upartial}{0}{UPM}{"40}
      \DeclareMathSymbol{\leqslant}{3}{AMSa}{"36}
      \DeclareMathSymbol{\geqslant}{3}{AMSa}{"3E}

      \let\geq=\geqslant 
    \fi
  \fi
\fi 

\ifCUPmtlplainloaded \else
  \ifAMStwofonts \else 
    \def\upi{\pi}
    \def\umu{\mu}
    \def\upartial{\partial}
  \fi
\fi

\title[Modelling the IR galaxy evolution]
 {Modelling the infrared galaxy evolution using a phenomenological approach}
\author[G. Lagache et al.]
  {G.~Lagache,$^1$
  H.~Dole,$^2$ J.-L.~Puget$^1$ \\
  $^1$Institut d'Astrophysique Spatiale, B\^at.  121, Universit\'e Paris XI, 91405 Orsay Cedex, France \\
  $^2$Steward Observatory, University of Arizona, 933 N Cherry Ave, Tucson, AZ, 85721, USA}
\date{Accepted 2002 August 14, Received 2002 June 6}
\pagerange{\pageref{firstpage}--\pageref{lastpage}}
\pubyear{2002}

\def\LaTeX{L\kern-.36em\raise.3ex\hbox{a}\kern-.15em
    T\kern-.1667em\lower.7ex\hbox{E}\kern-.125emX}

\begin{document}

\label{firstpage}

\maketitle

\begin{abstract}         
To characterise the cosmological evolution of the sources contributing to the infrared extragalactic
background, we have developped a phenomenological model that constrains in a simple way the galaxy 
luminosity function evolution with the redshift, and fits all the existing source counts and
redshift distributions, Cosmic Infrared Background intensity and
fluctuations observations, from the mid-infrared to the submillimetre range.
The model is based on template spectra of starburst and normal galaxies, and on the local
infrared luminosity function. Although the Cosmic Infrared Background can be modeled
with very different luminosity functions as long as the radiation production
with redshift is the right one, the number counts,
and the anisotropies of the unresolved background, imply that
the luminosity function must change dramatically with redshift, with a rapid
evolution of the high-luminosity sources (L$>$3 10$^{11}$ L$_{\odot})$
from z=0 to z=1 which then stay rather constant up to redshift 5. The derived evolution of the 
IR luminosity function may be linked to a bimodal star formation process, one 
associated with the quiescent and passive phase of the galaxy evolution and one associated with the starburst 
phase, trigerred by merging and interactions.
The latter dominates the infrared and submillimetre ouput energy of the Universe.
\\
The model is intended as a convenient
tool to plan further observations, as illustrated through  
predictions for {\it Herschel}, {\it Planck} and {\it ALMA} observations.
Our model predictions for given wavelengths, together with
some usefull routines, are available for general use.
\end{abstract}

\begin{keywords}
galaxies: evolution -- infrared: galaxies -- galaxies: starburst -- galaxies: general -- cosmology: observations
\end{keywords}

\section{Introduction}
The discovery of the Cosmic Infrared Background (CIB) (Puget et al. 1996; Fixsen et al. 1998; Hauser et al. 1998;
Schlegel et al. 1998; Lagache et al. 1999; Lagache et al. 2000; 
see Hauser \& Dwek 2001 for a review), together with
recent deep cosmological surveys in the infrared (IR)  and submillimetre (submm) has opened new perspectives
on our galaxy formation and evolution understanding. The surprisingly high amount of energy contained
in the CIB showed that it is crucial to probe its contributing galaxies 
to understand when and how the bulk of stars formed in the Universe.
Thanks to ISO (Kessler et al., 1996)
-- mainly at 15~$\mu$m with {\it ISOCAM} (Cesarsky et al. 1996), and 90 and 170~$\mu$m with
{\it ISOPHOT} (Lemke et al. 1996) -- 
and ground-based instruments 
-- {\it SCUBA} (Holland et al. 1998) and {\it MAMBO} (Bertoldi et al. 2000) at 850 and 1300 $\mu$m respectively --
deep cosmological surveys have been carried out. It is thus now possible, to various degrees, to resolve the CIB into
discrete sources (e.g.  Kawara et al. 1998; Barger et al. 1999; Elbaz et al. 1999; Carilli et al. 2000; 
Juvela et al. 2000; Linden-Vornle et al. 2000; Matsuhara et al. 2000; Bertoldi et al. 2001; Dole et al. 2001;
Elbaz et al. 2002; Scott et al. 2002). The striking result of these surveys
concerns the evolution of the IR and submm galaxy population.
The source counts are high when compared to 
no, or moderate, evolution models{\footnote{`No-evolution': 
the co-moving luminosity function remains equal to the local one at all
redshift}} for IR galaxies (Guiderdoni et al. 1998; Franceschini et al. 1998).
Therefore, it has been necessary to develop new models in the IR.
Very recently, several empirical approaches have been proposed
to model the high rate of evolution of
IR galaxies (e.g  Devriendt \& Guiderdoni 2000; Wang \& Biermann 2000; 
Charry \& Elbaz 2001; Franceschini et al. 2001; Malkan \& Stecker 2001;
Pearson 2001; Rowan-Robinson 2001; Takeuchi et al. 2001; Xu et al. 2001;
Balland et al. 2002; Wang 2002) which fit all existing source counts, redshift
distribution and CIB intensity and fluctuations, although often 
not all of them.
We present in this paper a new model whose preliminary results
were published by Dole et al. (2000). The originality of this model
was to empirically separate the evolution of the starburst galaxies
with respect to the normal galaxies, the current observations
implying a strong evolution of the bright part of the Luminosity
Function (LF). It was shown for the first time that only the starburst part
should evolve very rapidly 
between z=0 and z=2, the evolution rate being much higher in the IR than in any other 
wavelength domain. \\

We present here a more sophisticated and 
detailed version of the first model 
(Dole et al. 2000).
Our philosophy is to build the simplest model of the LF evolution, easily deliverable,
with the lowest number of parametres but accounting
for all observationnal data. We stress out the point that we include the CIB
fluctuations levels as measured by {\it ISOPHOT} (Lagache \& Puget
2000; Matsuhara et al. 2000) and {\it IRAS} (Miville-Desch\^enes et al. 2002)
as an extra constraint. Recent observations strongly suggest
that the bulk of the optical and IR extragalactic
background is made
of two distinct galaxy populations (see Sect. \ref{Nature}).
Therefore, we restrict our model in the wavelength
domain 10 - 1500~$\mu$m, our goal being to quantify
the evolution of IR galaxies. 
The paper is organised as follows: we first summarise
our present knowledge on the evolution of IR galaxies,
and on the nature of the sources contributing to the 
Extragalactic Background 
(Sect. 2). Then, we present the ingredients of the
model (Sect. 3). In Sect. 4, we discuss the galaxy
templates used in the model. We then present the parametrisation
of the local LF (Sect. 5). In Sect. 6 are
given the results of the model (evolution of the LF,
evolution of the luminosity density, number counts, z-distribution,
CIB intensity and fluctuations).
And finally the model is used for predictions for the
{\it Herschel} and {\it Planck} surveys (Sect. 7.2 and 7.3 respectively),
gives requirements for future large deep survey experiments (Sect. 7.4)
and predictions for {\it ALMA} observations (Sect. 7.5). 
A summary is given in Sect. 8.

\section{Our present knowledge}

\subsection{The strong evolution of IR galaxies: observationnal evidences}
There have been, in the last few years, strong observationnal evidences
indicating extremely high rates of evolution for IR galaxies.\\

First, galaxy evolution can
be observed through its imprint on the far-IR EB (Extragalactic Background).
Weakly constrained even as recently as 6 years ago, various observations
now measure or give upper/lower limits on the
background from the UV to the millimetre (mm) waveband (e.g. Dwek et al. 1998; Gispert et al.
2000; Hauser \& Dwek 2001). The data show the existence of a minimum between 3 and 10~$\mu$m
separating direct stellar radiation from the IR part due to
radiation re-emitted by dust. This re-emitted dust radiation
contains at least a comparable integrated power as the optical/near-IR, and
perhaps as much as 2.5 times more.  This ratio is much
larger than what is measured locally ($\sim$30$\%$). The CIB
is thus likely to be dominated by a population of strongly evolving redshifted
IR galaxies.  Since the long wavelength spectrum of the background is
significantly flatter than the spectrum of local star-forming galaxies,
it strongly constrains the far-IR radiation production rate history
(Gispert et al. 2000). The energy density must
increase by a factor larger than 10 between the present time and a
redshift $\sim 1$--2 and then stay rather constant at higher redshift
(till $\sim 3$), contrary to the ultraviolet radiation production rate
which decreases rapidly. \\

Second, several deep cosmological surveys at
15, 90, 170, 850 and $1300\,\mu$m have resolved a fraction of the CIB into
discrete sources. For all surveys, number counts indicate a very strong
cosmological evolution of IR galaxies, in total power radiated
but also in the shape of the LF. This is particularly obvious at submm
wavelengths where the EB is dominated by high luminosity galaxies
(see the {\it SCUBA} and {\it MAMBO} results).
The high rates of evolution exceed those measured
in other wavelength domain as well as those observed for
quasars and Active Galactic Nucleis (AGNs) .\\

Finally, high rates of evolution are suggested by the detection
of the CIB Poissonian fluctuations at a high level at 60 and
100 $\mu$m with {\it IRAS} (Miville-Dech\^enes et al., 2002)
and 170 $\mu$m with {\it ISOPHOT} (Lagache \& Puget 2000; Matsuhara et al. 2000). 
For example, Matsuhara et al. (2000) give the constraints on the galaxy number 
counts down to 35 mJy at 90 $\mu$m and 60 mJy at 170 $\mu$m, which indicate the 
existence of a strong evolution down to these fluxes in the counts. 

\subsection{\label{Nature} Sources making the Extragalactic Background}

\subsubsection{Optical versus IR and submm EB sources}
Recent observations show
that the bulk of the optical and IR EB is made by
two distinct galaxy populations (e.g. Aussel
et al. 1999). Therefore,
one of the key question is wether the dusty star forming
galaxies are recognizable from optical/near-IR
data alone.\\

In the local Universe, Sanders \& Mirabel (1996)
show that the bolometric luminosity of IR galaxies
is uncorrelated with optical spectra.
The color excess derived from the Balmer lines ratio
does not significantly depend on the IR luminosity,
IR color or optical spectral type (Veilleux et al. 1999).
In fact, neither the moderate strength of the heavily extinguished starburst
emission lines, nor their optical colors can distinguish them
from galaxies with more modest rates of star formation
(Elbaz et al. 1999; Trentham et al. 1999; Poggianti et al.
1999).
However, recently, Poggianti \& Wu (2000) and Poggianti et al. (2001)
show that the incidence of e(a)\footnote{Galaxies with strong Balmer absorption
lines and [OII] in emission (Poggianti \& Wu (2000) and references therein)} sources in the different
IR selected samples seems to suggest that the e(a) signature
might be capable of identifying from optical data alone
a population of heavily extinct starburst galaxies.
Reproducing the e(a) spectrum requires the youngest stellar
generations to be significantly more extinguished
by dust than older stellar populations, and
implies a strong ongoing star formation activity at a level higher
than in quiescent spirals.\\

At intermediate redshift ($0.3<z<1$), one of 
the main information on IR galaxies
comes from identifications of {\it ISOCAM} deep
fields. For example, 
Flores et al. (1999a,b) presented results of a deep survey 
of one of the CFRS fields at 6.75 and 15 $\mu$m. 
At 15 $\mu$m, most (71$\%$) of the sources with optical spectroscopy 
are classified as e(a) galaxies.
The FIR luminosities of the Flores et al. e(a) galaxies are between
5.7 10$^{10}$ and 2 10$^{12}$ L$_{\odot}$. This is the first confirmation of the IR
luminous nature of e(a) galaxies.
More recently, Rigopoulou et al. (2000) using VLT spectroscopy
found that optical {\it ISOCAM} counterparts in the HDF-south
are indistinguishable from the dusty luminous
e(a) galaxies.\\

All these studies at low and intermediate redshift seem to show that 
IR galaxies predominantly exhibit e(a) signatures in their optical spectra.
On the contrary, in optically selected surveys of field
galaxies, e(a) spectra are present but seems to be scarse
(less than 10$\%$, see for example Poggianti et al. 1999).
In fact evidence for high e(a) incidences
are found in merging or interacting
systems or active compact groups. A
complete study of the IR emission of e(a) galaxies
is still to be done.\\

At much higher redshift, there are other 
evidences that the optically selected samples
and bright IR samples (e.g. the SCUBA blank field sample)
are different. For example,
Chapman et al. (2000) have performed submm photometry for a sample
of Lyman Break galaxies whose UV properties imply high star formation
rates. They found that the integrated signal from their Lyman break
sample is undetected in the submm. This implies that
the population of Lyman break galaxies
does not constitute a large part of the 
detected blank field bright submm sources.\\

In conclusion, it is clear that the optical and IR EBs
are not dominated by the same population of galaxies.
Therefore, we restrict our model in the wavelength
domain 10-2000~$\mu$m, our goal being to quantify
the evolution of the IR galaxies. 

\subsubsection{The CIB sources}
In the HDF-N field (HST Deep Field North),
Aussel et al. (1999) and Elbaz et al. (1999)
find that 30 to 50 $\%$ of {\it ISOCAM} galaxies
are associated with optical sources showing 
complexe structures and morphological pecularities.
Moreover, Cohen et al. (2000) show,
in the HDF-N, that more than 90$\%$
of the faint {\it ISOCAM} sources are members of concentrations.
This shows that past or present interactions or merging play a large role
in trigerring the IR emission of galaxies.
All the studies of
{\it ISOCAM} field sources show that the bulk of the CIB
at 15~$\mu$m comes from galaxies which have bolometric
luminosities of about 10$^{11}$-10$^{12}$~L$_{\odot}$, high masses ($\sim$ 10$^{11}$~M$_{\odot}$)
and redshift between 0.5 and 2. They experience an intense stellar formation
(100~M$_{\odot}$/yr) which appears to be uncorrelated with
the faint blue galaxy population dominating
the optical counts at z$\sim$0.7 (Ellis 1997; Elbaz et al. 1999). 
In galaxy clusters all {\it ISOCAM} sources are found preferably
at the periphery where there is still some star formation (Biviano et al. 1999).\\

At longer wavelengths, source identifications
are much more difficult. Thus, 
characterising the nature of the galaxies is
a long time process. However, we already have some indications.
The {\it FIRBACK} survey at 170 $\mu$m 
detected about 200 galaxies (Dole et al. 2001) making less than 10$\%$ of
the CIB. Schematically {\it FIRBACK} sources comprise 2 populations: one cold and nearby
(L$\sim$ 10$^{9}$ -  10$^{11}$ L$_{\odot}$) and one cold or warm 
very luminous (L$\sim$ 10$^{12}$ L$_{\odot}$) with 
redshift lower than 1.2. The optical spectroscopy of the brightest
{\it FIRBACK} sources reveals an `{\it IRAS}-like' starburst nature (Dennefeld
et al. in prep; Patris et al. 2002) with a moderate star-formation rate
(10~M$_{\odot}$/yr). 
These results are very similar to those of
Kakazu et al. (2002) who found that 62$\%$ of the Lockman Hole
170 $\mu$m sources are at redshit below 0.3 with luminosities
lower than 10$^{12}$ L$_{\odot}$ (based on Arp220 spectrum),
the rest being mostly ultraluminous IR galaxies 
with redshift between 0.3 and 1. In their sample, the 170 $\mu$m
sources appear also to be powered primarily by star formation.\\
In the submm, the main 
indication comes from the {\it SCUBA} deep surveys (e.g. Hughes et al. 1998; Barger et al. 1999;
Eales et al. 1999; Scott et al. 2002).
These surveys suggest that faint 850~$\mu$m sources are mostly ultraluminous galaxies
at typical redshift between 1 and 4 (e.g. Eales et al. 2000).
{\it SCUBA} sources above 3 mJy account for 20-30 $\%$ of the EB at 850~$\mu$m.
The present data show that the bulk
of the submm EB is likely to reside in sources with 850 $\mu$m fluxes near 0.5 mJy.
Barger et al. (1999) estimate that the FIR luminosity of a characteristic
1 mJy source is in the range 4-5 10$^{11}$ L$_{\odot}$, which is also 
the typical luminosities of sources making the bulk
of the CIB at 15~$\mu$m. Moreover, as for the 15~$\mu$m sources, several groups
have suggested that the submm sources are associated with merger events
(e.g. Smail et al. 1998; Lilly et al. 1999).
All these results show that the integrated power of the LF at redshift
greater than $\sim$0.5 must be dominated by sources with luminosities of a few 10$^{11}$ L$_{\odot}$
while the local LF is dominated by sources with
luminosities of the order of 5 10$^{10}$ L$_{\odot}$.

\subsubsection{The AGN contribution to the IR ouput energy}
At low-z, Veilleux et al. (1999) and Lutz et al.(1999) show that
most of the IR sources are powered by starbursts. The AGN contribution appears
dominant only at very high luminosities (L$>$2 10$^{12}$ L$_{\odot}$). 
Also using {\it FIRBACK} nearby bright source spectroscopy, we find that
less than 10$\%$ of sources show AGN signs (Dennefeld et al.,
in prep; Patris et al. 2002). The same conclusions are reached by Kakazu et al. (2002).
At intermediate z, from optical and X-rays studies of 
{\it ISOCAM} sources making the bulk of the CIB, several groups show
that 15 $\mu$m sources are mostly starburst galaxies (e.g. Fadda et al. 2002). 
At much higher z, the main indication comes
from X-ray observations of {\it SCUBA} sources  
(Barger et al. 2001; Fabian et al. 2000; Hornschemeier 
et al. 2000; Severgnini et al. 2000). All these observations
of submm galaxies in the X-rays
are consistent with starburst-dominated emission.
However, recently, Page et al. (2002) present a result of {\it SCUBA}
observations of eight X-ray absorbed active galactic nuclei
from z=1 to z=2.8
and find, for half of them, a 850 $\mu$m submm counterpart.
Nevertheless, the high 850 $\mu$m fluxes (greater than 5.9 mJy)
suggest that these sources are hyperluminous galaxies.
Such galaxies do not dominate the IR output of
the Universe. \\

Considering the whole CIB energy budget, and
based on the assumptions that 10$\%$ of the mass
accreting into black hole is turned into energy and that the black hole
masses measured in the HDF (Ford et al. 1998) are typical of galaxies,
the AGN background energy would be 
at the order of 10$\%$ of that from stars
(Eales et al. 1999). These calculations are highly uncertain but are supported
by the work of Almaini et al. (1999). And recently, more direct evidences have been obtained.
For example, Severgnini et al. (2000)
show that the 2-10 keV sources making at least 75$\%$ of the X-ray background
in this band contribute for less than 7$\%$ to the submm
background.\\

On the modelisation side, several groups (e.g. Xu et al. 2001; Rowan-Robinson
2001) show that the starburst galaxy evolves much more
rapidly than the AGN-dominated sources, making the AGN contribution
to the CIB relatively small.\\

In conclusion, it is clear that AGNs do not dominate the IR output energy
of the Universe. Therefore, we will consider hereafter that IR galaxies are 
mostly powered by star formation and we use Spectral Energy Distributions (SED)
typical of these star-forming galaxies. The differences in SED for the small fraction
of AGN dominated IR galaxies would not affect significantly the results of our model
which is built with only two galaxy populations (`normal' and starburst) defined by their SEDs.

\section{Ingredients of the model}
At a given wavelength $\lambda$=$\lambda_0$, the flux $S_{\nu}$ of a source at redshift z, 
as a function of the rest-frame luminosity $L_{\nu}$
(in W/Hz) can be written:
\begin{equation}
S_{\nu} (L, z, \lambda=\lambda_0)=  \frac{(1+z)  \times K(L,z) \times L_{\nu}(L, \lambda = \lambda_0)}{4 \pi  D_L^2 } 
\end{equation}
where  D$_L$ is the luminosity distance and
K(L,z) is the K-correction factor defined as :
\begin{equation}
K(L,z) = \frac{L_{\nu(1+z)}}{L_{\nu(z=0)}}
\end{equation}
This correction is specific of the spectrum of the considered
population at given L and z.
In practice, the rest-frame luminosity L$_{\nu}$ is convolved by the band pass
filter centred on $\lambda$=$\lambda_0$.
The number of sources per solid angle and redshift interval is:
\begin{equation}
\frac{dN}{dzdLogL} (L, z)= N_0(L,z) \times (1+z)^3 \times \frac{dV}{dz}
\end{equation}
where dV/dz is the differential volume element fixed by the cosmology and N$_0$ is the number of sources per 
unit volume and luminosity interval
as a function of redshift. N$_0$ is given by the LF.\\
The differential counts at a given flux S and wavelength $\lambda$=$\lambda_0$ write as:
\begin{equation}
\frac{dN}{dS}= \int_{L} \int_{z} \frac{dN}{dzdLogL}(L,z) \times \frac{dz}{dS}(L,z) \times dLogL$$
\end{equation}
We then obtain the integral counts:
\begin{equation}
N(>S) = \int \frac{dN}{dS} dS
\end{equation}
the CIB intensity produced by all sources with S$<$S$_{max}$:
\begin{equation}
I_{CIB}= \int_{0}^{S_{max}} S \frac{dN}{dS} dS
\end{equation}
and the CIB intensity fluctuations (the shot noise) from sources
below a given detection limit S$_0$ (that could correspond to 
either the confusion or the sensitivity limit) measured by the level
of the white noise power spectrum:
\begin{equation}
P_{fluc} = \int_{0}^{S_0} S^2 \frac{dN}{dS} dS \quad \mathrm{Jy^{2}/sr}
\end{equation}

We build the simplest model with the lowest number of parametres and ingredients
which fits all the observations.
We first fix the cosmology ($\Omega_{\Lambda}$=0.7, $\Omega_{0}$=0.3 and h=0.65)
from the combination of the most recent CMB anisotropy measurements (e.g. de Bernardis et al. 2002) 
the distance-luminosity relation of type Ia supernovae (e.g. Perlmutter et al. 1999)
and the measure of galaxy distances using Cepheids (e.g. Freedman et al. 2001).
We then construct the 'normal' and starburst template spectra : at each population and luminosity 
is associated one spectrum (see Sect. \ref{template}). 
We finally search for the best evolution of the LF (Eq. 3) that reproduces the number counts
(Eq. 4-5), the CIB and its fluctuations (Eq. 6 and 7 respectively),
assuming that the LF evolution is represented by the independent evolution of the two populations.
A rather remarquable result, as will be seen later, is that two populations
only can fit all the data.

\section{Building galaxy templates}
\label{template}

\begin{figure}
\begin{center}
\epsfxsize=10.cm
\epsfbox{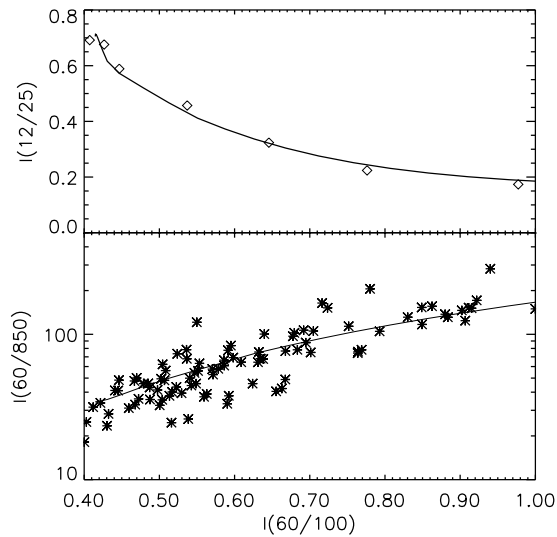}
\caption{\label{color_diag} Color diagrams from the model (continuous lines), compared with Soifer \& Neugeubauer 1991
(diamonds) and Dunne et al. 2000 (stars) data points.}
\end{center}
\end{figure}

\subsection{Starburst galaxies}
\begin{figure}
\begin{center}
\epsfxsize=9cm
\epsfbox{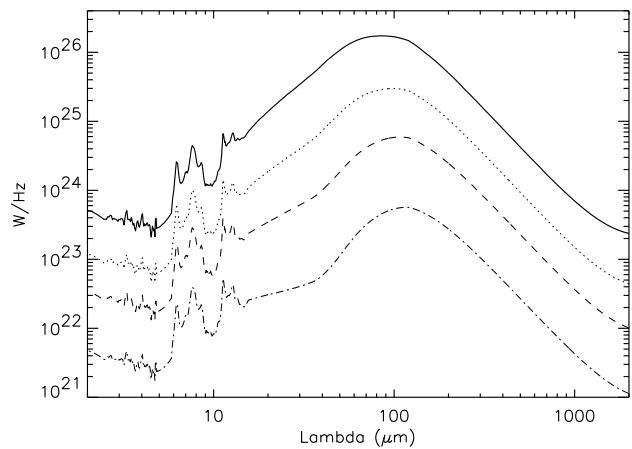}
\caption{\label{avg_spectra} Starburst model spectra for different luminosities: L=3. 10$^{12}$ L$_{\odot}$
(continuous line), L=5 10$^{11}$ L$_{\odot}$ (dotted line),  L=10$^{11}$ L$_{\odot}$ (dashed line) and
 L=10$^{10}$ L$_{\odot}$ (dotted-dashed line)}
\end{center}
\end{figure}

The luminous IR starburst galaxies emit more than 95$\%$ of their energy
in the far-IR. Spectra of such galaxies have been modeled by Maffei (1994),
using the D\'esert et al. (1990) dust emission model and
the observationnal correlation for the {\it IRAS} Bright Galaxy Sample of the flux ratio
12/60, 25/60 and 60/100 with the IR luminosity (Soifer \& Neugeubauer, 1991).
We start from this model and modify it slightly
to better take into account recent 
observationnal constraints.
The significant improvements are the following:
\begin{itemize}
\item First we replace the D\'esert et al. (1990) PAHs template
by the Dale et al. (2001) one, keeping the same amount of energy in
the mid-IR. 
\item We slighly modify the spectral shape in the near and mid-IR:
we increase the PAH and Very Small Grains proportions (by a factor
$\sim$2) but add some extinction, slowly increasing with the luminosity. 
\item Finally, we broaden the far-IR peak (also a continuous
change with the luminosity) and flattens slightly the long
wavelength spectrum (Dunne al. 2000; Dunne \& Eales 2001; Klaas et al. 2001).
\end{itemize}
Fig. \ref{color_diag} shows the (12/25, 60/100) and (60/850, 60/100) 
color diagrams 
compared with Soifer \& Neugeubauer (1991)
and Dunne et al. (2000) observations.
We see a very good agreement
between the templates and observations. 
The average luminosity spectra sequence is shown on Fig. \ref{avg_spectra}. 
For the `normal' starbursting galaxies (L$_{IR}<$10$^{11}$ L$_{\odot}$), we also check that the 7/15
versus 60/100 diagram was in good agreement
with Dale et al. (2001). \\

Such a representation of galaxy SEDs that assigns only one spectrum
per luminosity does not take into account the dispersion of the colors (e.g. the 60/100 ratio)
observed for a given luminosity (as in  Xu et al. 2001 or Chapman et al. 2002a).
However, we do not have enough measured colors to do a statistical analysis of their variations
with the luminosities. 
The only colors measured for a large sample are the IRAS
colors (principally the 60/100). Using only the 60/100 color as a tracer
of the dispersion for a given luminosity, may not improve 
the representation as for example two very different long-wavelength spectra can have the same 
60/100 color (e.g. NGC 7821 and NGC 0549, Stickel et al. 2000). 

\subsection{Normal spiral galaxies}
For the `normal' galaxies (i.e. standard IR counterparts of spiral
galaxies with more than half of their energy ouput in the optical), we take an unique spectrum derived mainly from 
the {\it ISOPHOT} serendipity survey (Stickel et al. 2000)
and the nearby {\it FIRBACK}
galaxy SEDs, together with longer wavelength data from
Dunne et al. (2000) and Dunne \& Eales (2001).\\

\begin{figure}
\begin{center}
\epsfxsize=7.5cm
\epsfysize=7.5cm
\epsfbox{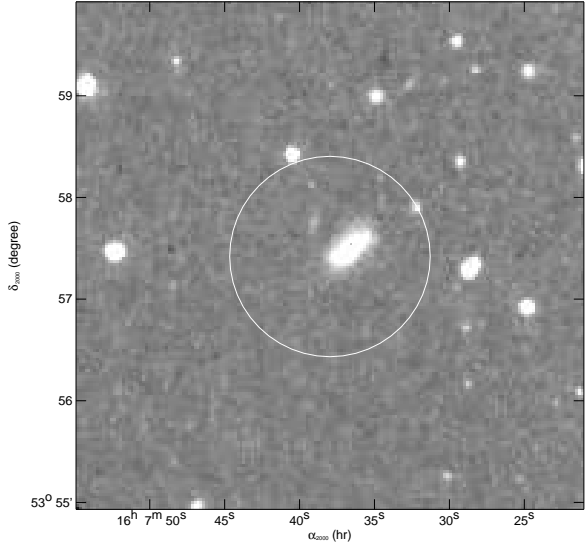}
\caption{\label{FIRBACK_cold} Example of association of one bright 
{\it FIRBACK} 170 $\mu$m source with a ``cold'' spiral galaxy
(the optical image is from the DSS; the white circle
corresponds to the 170 $\mu$m error position). 
IR color ratios for this source are I$_{60}$/I$_{100}$=0.6, I$_{170}$/I$_{100}$=1.4 and I$_{170}$/I$_{15}$=51}
\end{center}
\end{figure}

\begin{figure}
\begin{center}
\epsfxsize=9cm
\epsfbox{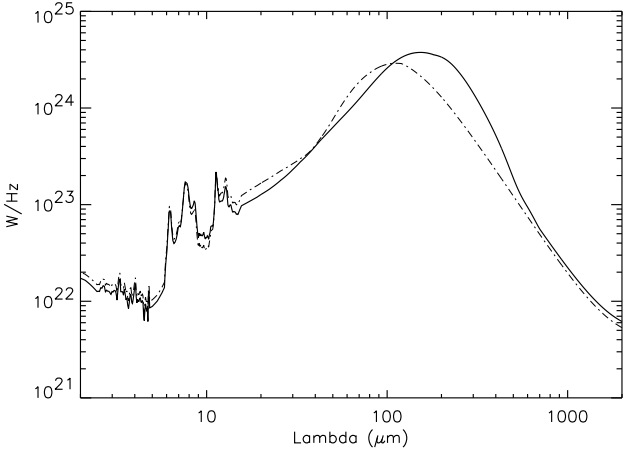}
\caption{\label{cold_spectrum} Template spectrum for the normal galaxies (continuous line),
compared with the template spectrum for the starburst galaxies (dash-dot line), for the same luminosity L=5 10$^{10}$ L$_{\odot}$}
\end{center}
\end{figure}

The {\it ISOPHOT} serendipity survey has revealed a population 
of nearby cold galaxies
(Stickel et al. 1998, 2000), under-represented
in the 60 $\mu$m {\it IRAS} sample. 
The distribution of I$_{170}$/I$_{100}$
flux ratio shows that about half of the galaxies have a flux ratio
between 1 and 1.5, indicating that the FIR spectra are mostly flat
between 100 and 200 $\mu$m. Very few show a 
downward trend in this wavelength
range, this trend being typical of warm starburst galaxies (see Fig. \ref{avg_spectra}).
Most important is the large fraction of sources (more than 40 $\%$)
which have I$_{170}$/I$_{100} >$1.5, indicating
a rising spectrum beyond 100 $\mu$m similar to that
seen for example in the Milky Way galactic ridge, a property
known for more than 20 years from early balloon-borne measurements
(Serra et al. 1978, Silverberg et al. 1978).\\
In the {\it FIRBACK} N1 field (Dennefeld et al., in prep), the brightest sources 
show mean IR color similar to that of Stickel
et al. (2000):  I$_{170}$/I$_{100} \sim $ 1.3
and I$_{60}$/I$_{100} \sim $0.47. 
These objects are often associated with bright optical spiral
galaxies (e.g. Fig. \ref{FIRBACK_cold}) with also
typical 15 $\mu$m fluxes such as I$_{170}$/I$_{15} \sim $42. 
Finally recent observations at 450 $\mu$m 
(Dunne \& Eales 2000) reveal also the presence, in normal galaxies, of a colder component
than previously thought.\\
Therefore, we take for the normal galaxie the ``cold'' template
presented in Fig \ref{cold_spectrum}. This template have mean IR color
of I$_{170}$/I$_{100}$= 1.42, I$_{60}$/I$_{100}$= 0.35,
I$_{170}$/I$_{15}$=45 and I$_{100}$/I$_{850}$= 77.
For the Mid-IR part of the template, we use the spectral signature 
that applies to the majority of star-forming galaxy presented
in Helou (2000).
More observations around the maximum intensity ($\sim$ 100-200 $\mu$m)
and in the submm
are needed if we want to refine the template and describe its variations
with the luminosity.\\

Note that we do not make any evolution of the 'normal' and starburst template spectra with the redshift.
As shown in Chapman et al. (2002a), the available data for high redshift, far-IR galaxies do not
reveal evidence for any strong evolution in the characteristic temperature of the color
distribution over 0$<z<$3. We can also test this hypothesis of 'no-evolution' using
the two high-redshift sources N1-040 and N1-064 detected in the FIRBACK N1 field (Chapman et al. 2002b).
For these two sources, we do a blind search for both the luminosity and redshift using
our template spectra, based on a $\chi^2$ test. For N1-064
the 'normal' and starburst template gives the same $\chi^2$ with Log(L)=12.2 and z=0.5
and Log(L)=12.8 and z=1.05 respectively. The starburst template gives results 
in good agreement with Chapman et al. (2002b). For N1-040, the best $\chi^2$ is
obtained for the 'normal' template with Log(L) = 12.2 and z=0.45, which is in perfect
agreement with Chapman et al. (2002b). The template spectra give photometric redshifts
in good agreement with the spectroscopic redshifts, suggesting no strong evolution
of the galaxy SEDs over $0<z<1$.

\section{Parametrisation of the local luminosity function}
\begin{figure}
\begin{center}
\epsfxsize=9cm
\epsfxsize=9cm
\epsfbox{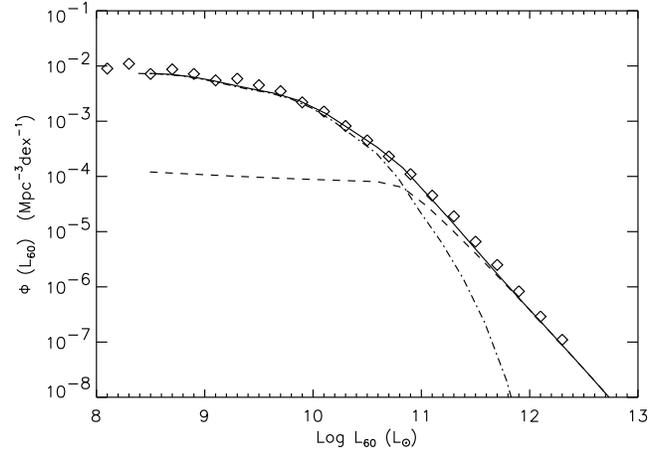}
\caption{\label{LF_60} Luminosity function at 60 $\mu$m (model: normal galaxies: dot-dash line, starburst galaxies:
dash line, total: continuous line) compared with Saunders et al. 1990 (squares).}
\end{center}
\end{figure}

\begin{figure}
\begin{center}
\epsfxsize=9cm
\epsfxsize=9cm
\epsfbox{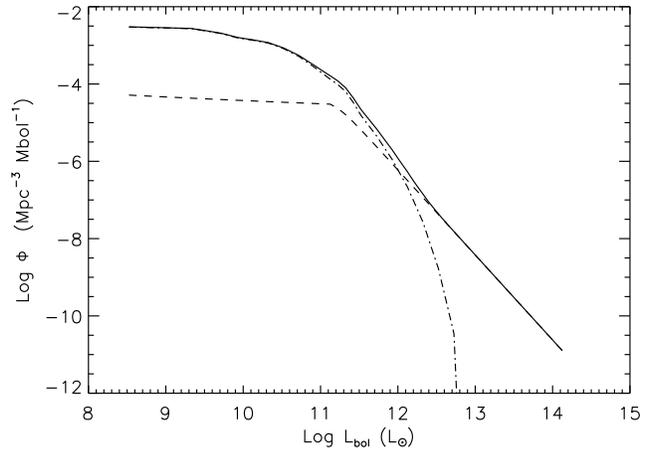}
\caption{\label{LF_bolo} Bolometric luminosity function at z=0 (normal galaxies: dot-dash, starburst galaxies:
dash and total: continuous)}
\end{center}
\end{figure}

A detailed comparison of the luminosity function of IR bright galaxies
with other classes of extragalactic objects has been done in Sanders \& Mirabel (1996).
The most striking results are (i)
the high luminosity tail of the IR galaxy LF is clearly in excess of what expected
from a Shechter function and (ii) at luminosities below $\sim$10$^{11}$ L$_{\odot}$, the majority
of optically selected objects are relatively weak far-IR emitters
(as shown by {\it IRAS}, {\it ISOPHOT} and {\it SCUBA} observations).
Accordingly, we decompose the IR LF in two parts:
(1) the `low-luminosity' part (dominated by normal galaxies)
that follows the shape of the optical LF and (2) the `high-luminosity' 
part dominated by starburst galaxies (note however that each part
covers the whole range of luminosities).\\

We start from the local LF at 60 $\mu$m from Saunders et al. (1990) (Fig. \ref{LF_60}).
We convert the 60 $\mu$m LF in a bolometric LF (L=1-1000 $\mu$m) using our
template spectra: at the 'low-luminosity' and 'high-luminosity' part, we assign the 'normal' 
and starburst template spectra respectively{\footnote{We have also
checked the validity of the long-wavelength part of our template spectra
by comparing the 850 $\mu$m LF with Dunne et al. (2000)}.
The bolometric LF (1-1000 $\mu$m) used in the model
is shown in Fig. \ref{LF_bolo}.\\

The parametrisation and redshift evolution of the two parts of the bolometric
LF is done using the following: 

\begin{itemize}
\item For the normal galaxies: 
\begin{itemize}
\item A LF at z=0 parametrised by an exponential with a cutoff in luminosity L$_{cutoff}$:
\begin{equation}
\Phi_{normal}(L) = \Phi_{bol}(L) \times exp  \left ( \frac{-L}{L_{cutoff}} \right ) 
\end{equation}
\item A weak number evolution
\end{itemize}
\item For the starburst galaxies:
\begin{itemize}
\item A LF described by:
\begin{equation}
\Phi_{SB}(L,z) = \Phi_{bol}(L=2.5 \quad 10^{11} L_{\odot}) 
\end{equation}
$$\times \left ( \frac{L}{2.5 \quad 10^{11}  L_{\odot}} \right )^{SB_{slope}} 
\times exp  \left ( - \frac{L_{knee}(z)}{L} \right )^3 $$ 
\item An evolution of the luminosity of the knee (L$_{knee}$) with the redshift
\item A constant slope of the LF at high luminosities (SB$_{slope}$), not redshift dependent
\item An evolution of the luminosity density with the redshift, $\varphi(z)$, that drives the LF evolution:
\begin{equation}
\Phi_{SB}(L,z) = \Phi_{SB}(L,z) \times \left ( \frac{\varphi(z)}{\varphi(z=0)} \right )
\end{equation}
\item An integrated energy at z=0 as observed: SB$_{norm}$=$\varphi(z=0)$ 
\end{itemize}
\end{itemize}
}

\section{Results and Discussions}

\subsection{LF evolution derived from observations}
\begin{figure*}
\begin{center}
\epsfxsize=9cm
\epsfbox{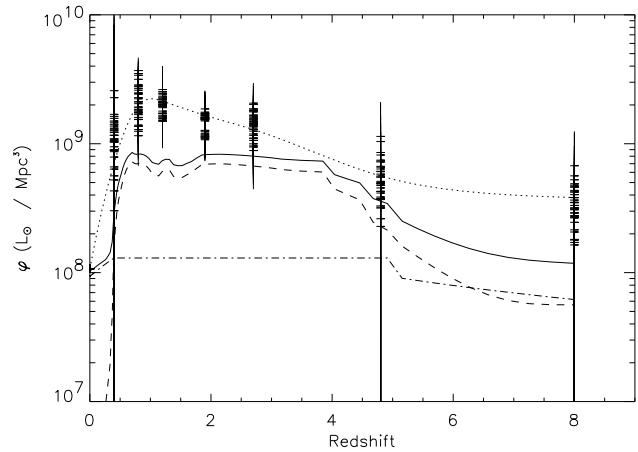}
\caption{\label{Lum_density} Co-moving luminosity density distribution
for the starburst galaxies (dash line), normal galaxies
(dot-dash line), and both normal and starburst galaxies (continuous line). Also shown for comparison
is the co-moving luminosity density distribution from all
cases of Gispert et al. 2001 (crosses with error bars), 
together with the best fit passing through all cases (dot line).}
\end{center}
\end{figure*}

Although the number of parametres is quite low, it is
too much time consuming to do a blind search through the whole
parametre space. We therefore search the best solution
for the parametres near values expected from direct observationnal
evidences:
\begin{itemize}
\item We fix the normal galaxy evolution (passive evolution) such as it nearly follows the number density evolution
of optical counts up to z=0.4: \\
$\Phi_{normal}(L,z) = \Phi_{normal}(L,z=0) \times (1+z)$. \\
We arbitrarily stop the evolution at z=0.4 and keep this population constant up to z=5 and then let 
the population decreases up to z=8 (see Fig. \ref{Lum_density})
\item An estimate of L$_{knee}$ is given by deep surveys at 15 and 850 $\mu$m: the bulk of the CIB
at 15 $\mu$m and 850 $\mu$m is made by galaxies with L$\sim$1-5 10$^{11}$
L$_{\odot}$ (e.g. Barger et al. 1999; Elbaz et al. 2002)
\item We have indications on the evolution of the luminosity density, $\varphi(z)$, for the starburst
galaxies, directly inferred from the CIB spectrum shape by Gispert et al. (2000) which
is used as a starting point to adjust the model.
\item We take SB$_{norm}$=10$^7$ L$_{\odot}$/Mpc$^3$ which is roughly the value observed
locally.
\item We take SB$_{slope}$= -2.2 (Kim \& Sanders, 1998)
\end{itemize}
The best evolution of the LF that reproduces
IR number counts, redshift distributions and CIB observations
is shown on Fig. \ref{LF_evol}. It is obtained with:
\begin{itemize}
\item L$_{cutoff}$ = 5 10$^{11}$ L$_{\odot}$
\item L$_{knee}$(z) = 8 10$^{10}$ $\times$ (1+z)$^3$ L$_{\odot}$ up to z=1.5 and then
 L$_{knee}$(z)= L$_{knee}$(z=1.5).
\end{itemize}
Moreover, to avoid a 'break' in the evolved luminosity function (as in Dole et al. 2000),
we modify the low-luminosity part of the starburst LF with the redshift (up to z=5) according to:
\begin{equation}
\Phi_{SB}(L,z) = \Phi_{SB}(L=L_{max},z) \times \left ( \frac{L_{max}}{L}\right )^{(1+z)^2}
\end{equation}
where $L_{max}$ is the luminosity corresponding to the maximum of $\Phi_{SB}(L,z)$.
\\

We see a very high rate of evolution
of the starburst part which peaks at z$\sim$0.7
(Fig. \ref{Lum_density}) and then
remains nearly constant up to z=4
(as shown for example by Charry \& Elbaz, 2001).
We compare on Fig. \ref{Lum_density} the co-moving luminosity density distribution
as derived from the model with the Gispert et al. (2000) one.
There is a good overall agreement. However, the model is systematically
lower than the Gispert et al. (2000)
determination for redshifts between 0.5 and 2.
This comes from the fact that the CIB values at 100 and 140~$\mu$m 
used in Gispert et al. (2000) were slightly overestimated (Renault et al. 2001)
leading in an overestimate of the luminosity density distribution at low redshift.\\

\subsection{Comparison model-observations}

\subsubsection{The number counts}
\begin{figure*}
\begin{center}
\epsfxsize=15cm
\epsfbox{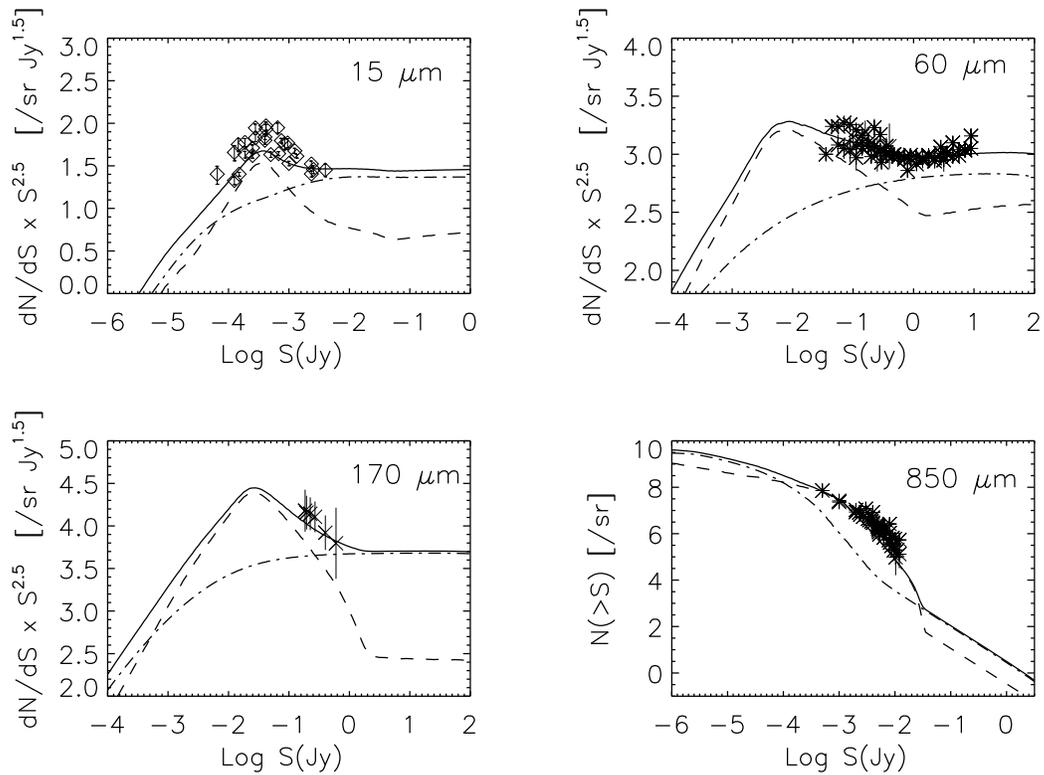}
\caption{\label{number_counts} Number counts at 15, 60, 170 and 850 $\mu$m (in Log)
together with the model predictions (starburst galaxies: dash line, normal galaxies:
dot-dash line, both normal and starburst galaxies: continuous line).
Data at 15 $\mu$m are from Elbaz et al. (1999), at 170 $\mu$m from Dole et al. (2001),
at 60 $\mu$m from Hacking \& Houck (1987), Gregorich et al. (1995), Bertin et al. (1997), 
Lonsdale et al. (1990), Saunders et al. (1990) and Rowan-Robinson et al. (1990) and at 850 $\mu$m from 
Smail et al. (1997), Hughes et al. (1998), Barger et al. (1999), Blain et al. (1999), 
Borys et al. (2001), Scott et al. (2002) and Webb et al. (2002).}
\end{center}
\end{figure*}

Fig. \ref{number_counts} shows the comparison of the 
number counts at 15~$\mu$m (Elbaz et al. 1999), 60~$\mu$m
(Hacking \& Houck 1987; Gregorich et al. 1995; Bertin et al. 1997; 
Lonsdale et al. 1990; Saunders et al. 1990; Rowan-Robinson et al. 1990), 
170 ~$\mu$m (Dole et al. 2001) and 850~$\mu$m
(Smail et al. 1997; Hughes et al. 1998; Barger et al. 1999; Blain et al. 1999; 
Borys et al. 2002; Scott et al. 2002; Webb et al. 2002)
with the observations.
We have a very good overall agreement (we also agree with the 90 $\mu$m
number counts of Serjeant et al. 2001 and Linden-Vornle et al. 2001). \\

\subsubsection{The redshift distributions: Need for a normal 'cold' population}

\begin{figure*}
\begin{center}
\epsfxsize=15cm
\epsfbox{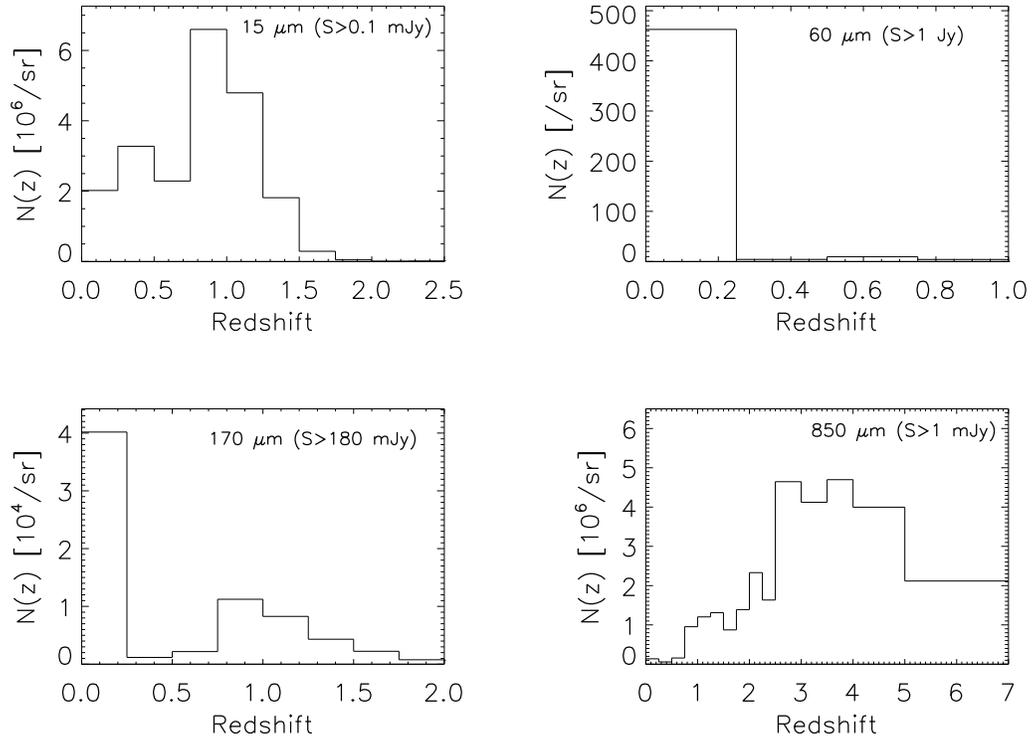}
\caption{\label{z-distrib} Predicted redshift distribution for resolved sources at 15, 60, 170 and 850
$\mu$m.}
\end{center}
\end{figure*}

On Fig. \ref{z-distrib} is shown the redshift distribution of resolved sources 
at 15, 60, 170 and 850 $\mu$m.
The 15 $\mu$m redshift distribution is in very good agreement with that observed by 
Flores et al. (1999b) and Aussel et al. (1999). At 850 $\mu$m, we predict 
that most of the detected sources are at $z>2.5$. At 170 $\mu$m, we predict that about
62$\%$ of sources with fluxes $S>$180 mJy (4$\sigma$ in Dole et al. 2001)
are at redshift below 0.25, the rest being mostly at 
redshift between 0.8 and 1.2. We know from the {\it FIRBACK} observations that the 
redshift distribution predicted by the model is very close to that observed
at 170 $\mu$m: it is clear from these observations that we have a bi-modal
z-distribution (Sajina et al. 2002; Dennefeld et al. in prep). 
Moreover, very recently, Kakazu et al. (2002) published
first results from optical spectroscopy of 170~$\mu$m
Lockman Hole sources. They find that 62$\%$ of sources
are at z$<$0.3 with IR luminosities (derived using Arp220 SED) 
lower than 10$^{12}$ L$_{\odot}$,
the rest being at redshift between 0.3 and 1, which is in very good
agreement with the model prediction. It is very important to note that the
agreement between the model and observations can only be obtained 
with the local `cold' population.
A model containing only starburst-like template spectra
and `warm' normal galaxies
overpredicts by a large factor the peak at z$\sim$1
(as in Charry \& Elbaz 2001 for example). 

\subsubsection{The CIB and its anisotropies}

\begin{figure*}
\begin{center}
\epsfxsize=15cm
\epsfbox{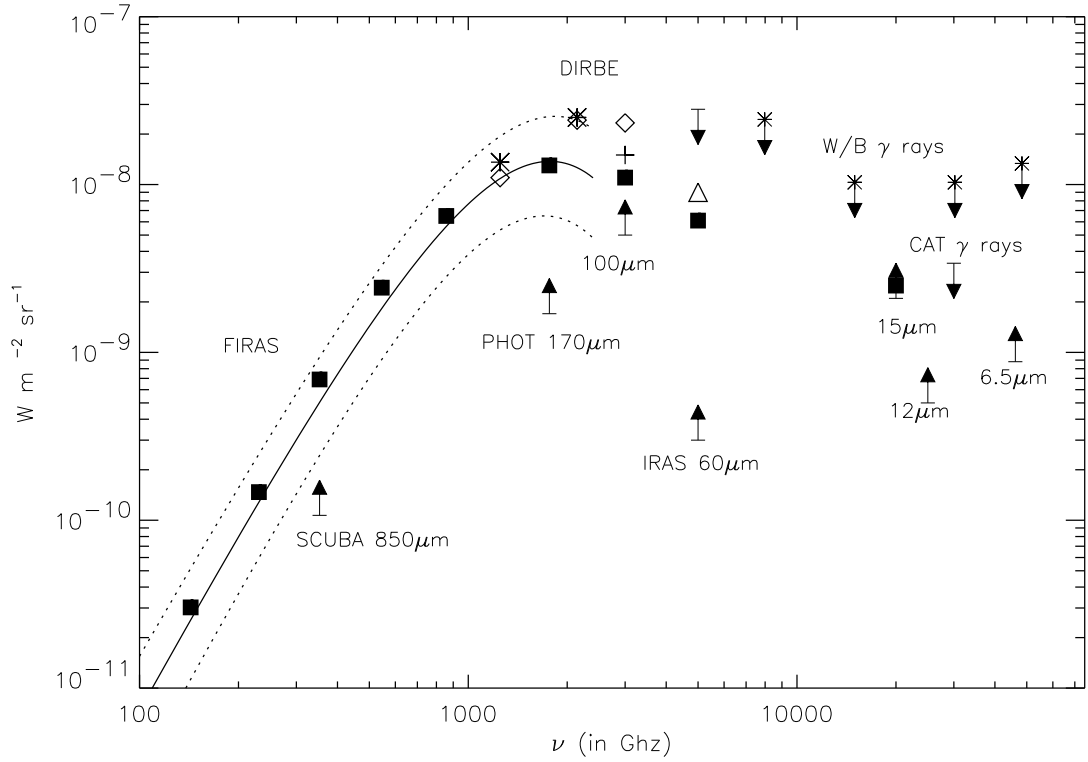}
\caption{\label{CIB_fig} Cosmic Background from mid-IR to the
millimetre wavelength. The 6.5
(D\'esert, private communication), 12 (Clements et al. 1999) and 15~$\mu$m (Elbaz et al 2002)
lower limits come from ISOCAM number counts;
the upper limit ``CAT'' is from Renault et al. (2001) and the cross upper limits W/B are from
Biller et al. (1998). At longer wavelength, we have the 60 $\mu$m estimate from Miville-Desch\^enes et al. 2002
($\triangle$), the upper limit from Finkbeiner et al. (2000), and the lower limit from number 
counts at 60 $\mu$m (Lonsdale et al. 1990);
at 140 and 240 $\mu$m are displayed the  
Lagache et al. 2000 ($\diamond$) and Hauser et al. 1998 ($\star$)
DIRBE values; at 100 $\mu$m is given the lower limit from Dwek et al. (1998) together with the estimates
from Renault et al. 2001 ($+$) and the determination of
Lagache et al. 2000 ($\diamond$); at 170 (Puget et al. 1999) and 850 $\mu$m (Barger et al. 1999) are
lower limits from number counts. The analytic form of the CIB at the FIRAS
wavelengths is from Fixsen et al. (1998). The CIB derived from the model at selected 
wavelengths is given by filled squares.}
\end{center}
\end{figure*}

The predicted CIB intensity at specific wavelengths, together with the
comparison with present observations are presented in Table \ref{CIB-tbl}
and in Fig. \ref{CIB_fig}.
We have a very good agreement with the estimates at 60 $\mu$m
(Miville-Desch\^enes et al. 2002), 100 $\mu$m (Renault et al. 2001),
and 170 $\mu$m (Kiss et al. 2001) and the {\it FIRAS} determinations 
(Fixsen et al. 1998; Lagache et al. 2000). We are also in good
agreement with the lower limit derived from 15 $\mu$m counts (Elbaz et al. 2002),
combined with the upper limit deduced from high energy gamma-ray emission
of the active galactic nucleus Mkn 501 (Renault et al. 2001).
According to the model, sources above 1 mJy at 850 $\mu$m contribute for
about 30$\%$ of the CIB. At 15 $\mu$m, with the deepest {\it ISOCAM} observations
(Altieri et al. 1999; Aussel et al. 1999; Metcalfe 2000)
about 70 $\%$ of the CIB has been resolved into individual sources.\\

Finally, we compare the model predictions for the CIB fluctuations
with the present observations at 60, 90, 100 and 170~$\mu$m (Table \ref{Fluc-tbl}).
For the comparison we need to remove the contribution
of the brightest sources that make the bulk of the fluctuations
(sources with fluxes $S>S_{max}$). When S$_{max}$ is quite high
(at the order of 500 mJy - 2 Jy), the fluctuations are dominated
by the strongest sources, making the result very dependent on the accuracy
of the evaluation of S$_{max}$. 
This is why the comparison observations/model
is very difficult{\footnote{
We cannot compare the model predictions with the Kiss et al. (2001)
results since we have no information on S$_{max}$}}. 
For S$_{max}$ around 50-150 mJy, the fluctuations
are dominated by the faint and numerous sources that 
dominate the CIB and the values do not depend critically
on the exact value of S$_{max}$.
We see from Table \ref{Fluc-tbl} that although for some
observations the model can be lower or greater by factor
1.5 in amplitude, we have an overall very good
agreement. We stress out that reproducing the CIB
fluctuations gives strong constraints on the
LF evolution. For example, evolution
such that we better reproduce the 850 $\mu$m counts
gives too high CIB fluctuations. Future observations
with better accuracy will show if these minor discrepancies
disappear or are indicative that the bright submm counts are 
overestimated due to a high fraction of gravitationally lensed
sources (Perrotta et al. 2002), 
or are simply indicative that the present
phenomenological model is too simple !
The level of the predicted CIB fluctuations for dedicated
experiments (as for example {\it Boomerang} or {\it Maxima}), with respect to the cirrus 
confusion noise and instrumental noise will be discussed in detail in Piat et al. (in prep).
\\
 
\begin{figure*}
\begin{center}
\epsfxsize=10cm
\epsfbox{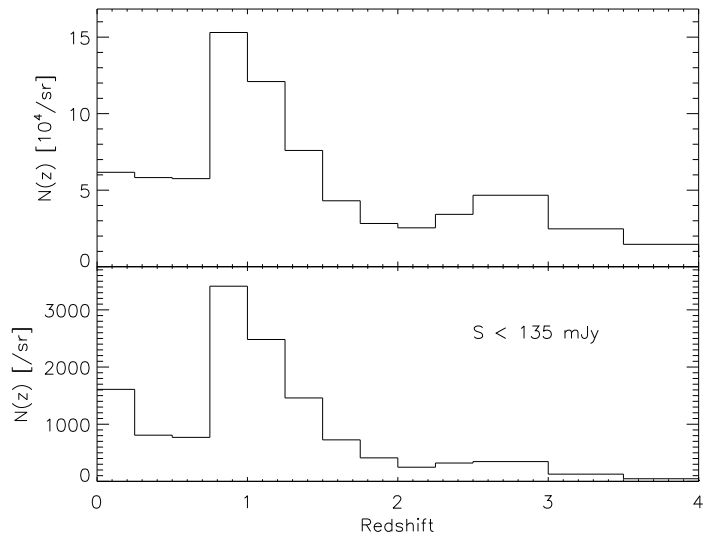}
\caption{\label{CIB_fluc_170} Redshift distribution of sources making
the CIB intensity and its fluctuations (S$<$135 mJy) at 170 $\mu$m.}
\end{center}
\end{figure*}

At 170 $\mu$m, it is clear from Fig. \ref{CIB_fluc_170} 
that the redshift distributions of sources that are 
making the CIB and those making the bulk of
the fluctuations are similar. 
The fluctuations are not dominated by bright
sources just below the detection threshold
but by numerous sources at higher redshift.
Thus, in this case, studying the CIB fluctuations gives
strong constraints on the CIB source population. This is true in the
whole submm-mm range.
The population of galaxies, where the CIB peaks, will not
be accessible by direct detection in the coming years. 
For example, {\it SIRTF} will resolve about 20$\%$ of
the background at 160 $\mu$m (Dole et al. 2002), 
{\it PACS} about 50 $\%$ at 170 $\mu$m (Sect. 7.2.2)
and {\it SPIRE} less than 10$\%$ at 350 $\mu$m (Sect. 7.2.1).\\

In conclusion, we have seen that our model gives number counts, 
redshift distributions, CIB intensity and fluctuations
that reproduce all the present observations.
It can be now used for future experiment predictions,
in particular for {\it Herschel}, {\it Planck} and {\it ALMA} observations.
For {\it SIRTF}, a complete and more detailed study, including
simulations and a detailed discussion on the confusion, is done
in Dole et al. (2002).

\section{Predictions for future experiments}

\subsection{The confusion noise from extragalactic sources}
The confusion noise{\footnote{We only consider the confusion noise
due to extragalactic sources since, in the high galactic latitude cosmological fields,
the cirrus confusion noise is negligible.}} is usually defined as the fluctuations of the background sky 
brightness below which sources cannot be detected individually. 
These fluctuations are caused by intrinsically discrete extragalactic sources. 
In the far-IR, submm and mm wavelengths, due to the limited size
of the telescopes compared to the wavelength, the confusion noise
play an important role in the total noise bugdet. In fact, the confusion 
noise is often greater than the instrumental noise, and is thus limiting
severely the surveys depth. 
The total variance $\sigma^2$ of a measurement within a beam due
to extragalactic sources with fluxes less than S$_{lim}$
is given by:
\begin{equation}
\label{eq_sigma}
\sigma^2 =  \int f^2(\theta, \phi) d\theta d\phi \int_{0}^{S_{lim}} S^2 \frac{dN}{dS} dS
\end{equation}
where $f(\theta, \phi)$ is the two-dimensional beam profile (in steradians), S the flux in Jy,
and $\frac{dN}{dS}$ the differential number counts in Jy$^{-1}$ sr$^{-1}$. We call S$_{lim}$
the confusion limit.\\

The confusion noise can be determined using
two criteria: the so-called photometric and source density
criteria (see Dole et al. 2002 for a full description).
The photometric criterion is related to the quality of the photometry
of detected sources, the flux measured  near S$_{lim}$ being severely affected by
the presence of fainter sources in the beam. It is defined by the implicit equation: 
\begin{equation}
S_{lim} = q_{phot} \times \sigma(S_{lim})
\end{equation}
where q$_{phot}$ measures the photometric
accuracy and is usually taken between 3 and 5.\\

The source density criterion is related to the completness of detected sources
above $S_{lim}$ directly related to
the probability to lose sources too close to each other
to be separated.
There is a threshold above which the density of sources above
$S_{lim}$
is such that a significant fraction of the sources is lost (it is impossible to 
separate the individual sources anymore). For a given source density 
(with a Poissonian distribution) N($>$S), corresponding to
a flux limit S$_{lim}$, the probability P to
have the nearest source with flux greater than S$_{lim}$
located closer than the distance $\theta_{min}$
is:
\begin{equation}
\label{eq_P}
P(<\theta_{min}) = 1 - e^{-\pi N \theta^2_{min}}
\end{equation}
$\theta_{min}$ is the distance below which sources cannot
be separated and is a function of the beam profile.
If we note $\theta_{FW}$, the FWHM of the beam profile,
$\theta_{min}$ can conveniently be expressed in unity of
$\theta_{FW}$, $\theta_{min}$ = k$\times$ $\theta_{FW}$.
As an illustration, simulations of source extraction at the {\it MIPS/SIRTF}
wavelengths show that k=0.8 should be achievable (Dole et al. 2002).
Therefore, in the following, we fix k=0.8.
We choose the maximum acceptable probability of not being
able to separate the nearest source 
P$_{max}$ = 0.1. In this case, the source density is equal
to 1 / 17.3$\Omega$ (Table \ref{N_omega}) and
the corresponding most probable distance is about 1.7$\times$$\theta_{FW}$.\\
Using Eq. \ref{eq_P}, we can derive N($>$S) and thus
find the corresponding S$_{lim}$. Then, using Eq. \ref{eq_sigma}, we compute $\sigma$.
The source density criterion leads to a value of an equivalent
q$_{density}$ = $\frac{S_{lim}}{\sigma}$. If q$_{density}$ is greater
than standard values of q$_{phot}$ (3 to 5), then the confusion noise
is given by the source density criterion. If not, then the
photometric criterion has to be used to derive the
confusion noise. The classical confusion limit of 1 source per 30 beams
corresponds to k=1 and P$\sim$0.1. Nevertheless it can still lead to mediocre
photometry for very steep logN-logS.\\

\begin{table}
\begin{center}
\caption{Number of beams per sources for different probability P to
have the nearest source 
located closer than the distance $\theta_{min}$~=~k$\times$ $\theta_{FW}$
with $\theta_{FW}^2$~$\sim$~$\Omega$/1.1 (valid for both Gaussian and Airy disk beams). }
\label{N_omega}
\begin{tabular}{|c|c|c|c|} \hline 
      & P = 0.05 & P = 0.1 & P = 0.15 \\ \hline
k=0.8 & 35.6 & 17.3 & 11.2 \\ \hline 
k=0.9 & 45.1 & 22.0 & 14.2 \\ \hline
k=1.0 & 55.7 & 27.1 & 17.6 \\ \hline
\end{tabular}\\

\end{center}
\end{table}

The transition between the photometric and source density criterion is around
200 $\mu$m, depending on telescope diameters. 
For example for {\it SIRTF}, at 24 and 70 $\mu$m,
the source density criterion gives 
q$_{density} \sim$7 (Dole et al. 2002). In this case, 
the source density criterion has to be used 
to derive the source confusion and it leads
to a very good photometric quality. On the contrary, at the {\it Planck} wavelengths,
the source density criterion gives q$_{density}\sim$1,
leading the source detection limited by the photometric quality. 
Furthermore, this illustrates the limits of doing high signal-to-noise observations
`to beat the confusion'.\\

\begin{figure*}
\begin{center}
\epsfxsize=10cm
\epsfbox{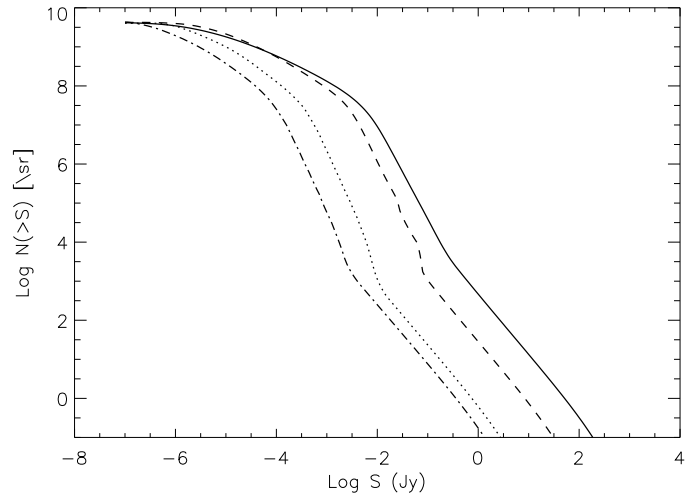}
\caption{\label{counts1} Predicted Number counts at 350 (continuous line), 550 $\mu$m (dashed line), 1300 $\mu$m
(dotted line) and 2097 $\mu$m (dotted-dashed line).}
\end{center}
\end{figure*}

In the following, we derive the confusion
noise for future long wavelength dedicated surveys. We assume that the sources 
are randomly distributed on the sky. The effect 
of the clustering on the confusion noise will
be investigated by Blaizot et al. (in prep)
using the hybrid model described in Guiderdoni et al. (2001) and Hatton et
al. (2002).
The number counts at long wavelengths derived from the model and usefull
for the following predictions are shown on Fig. \ref{counts1}.

\subsection{The {\it Herschel} deep surveys}
The `{\it Herschel} Space Observatory' (Pilbratt, 2001) is the fourth cornerstone mission in the European Space Agency 
science programme. It will perform imaging photometry and spectroscopy in the far-IR 
and submm part of the spectrum, covering approximately the 60-670~$\mu$m range. 
{\it Herschel} will carry a 3.5 metre diameter passively cooled telescope. 
Three instruments share the focal plane:  two cameras/medium resolution spectrometres, {\it PACS} and
{\it SPIRE}, and a very high resolution heterodyne spectrometre, {\it HIFI}.\\

In Table  \ref{conf_H} are shown the confusion noises and limits at {\it PACS} and {\it SPIRE}
wavelengths using the photometric
and source density criteria.
At the {\it PACS} wavelengths, the source density criterion leads
to confusion limits 2 to 6 times higher than the photometric
criterion with q$_{phot}$=5. For {\it SPIRE}, at 250~$\mu$m,
the limits obtained using the two criteria are very close
and then, at longer wavelengths, the photometric
quality constrains the confusion limit.\\

Extragalactic surveys will be conducted
by both the {\it PACS} and {\it SPIRE} instruments. Up to now, the trade-off between
large-area, shallow versus small-area, deep/ultradeep
observations has not been finalised. In the following two sections
are examples of extragalactic surveys that could be done. 

\subsubsection{The {\it SPIRE} surveys}
\begin{figure*}
\begin{center}
\epsfxsize=10cm
\epsfbox{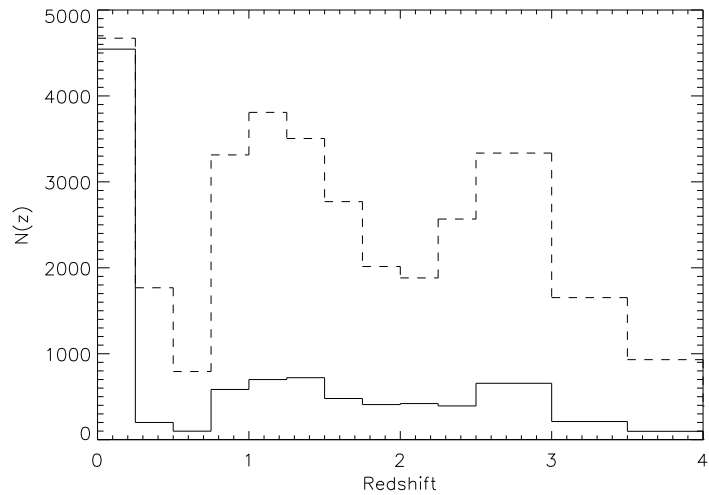}
\caption{\label{z-350} {\it SPIRE} 350 $\mu$m redshift distribution for sources detected
in the very-large area survey (continuous line, S$>$103.9 mJy, multiplied by 2) and the confusion-limited
survey (dashed line, S$>$27.1 mJy).}
\end{center}
\end{figure*}

For large-area scan-mapping observations, the current
estimates of time needed to map
1 square degree to an instrumental noise level of 3 mJy (1$\sigma_{inst}$)
is about 1.7, 2 and 2.1 days at 250, 350 and 550 $\mu$m 
respectively (Matt Griffin, private communication).
Three kinds of surveys could be done with {\it SPIRE} (Table \ref{Sens_SPIRE}):
\begin{enumerate}
\item A very large-area (2 approximatly $14^o\times14^o$, i.e. $\sim$ 400 square degrees)
survey at the noise level of 5$\sigma_{inst}$=100 mJy at 350 $\mu$m (92 mJy
at 250 $\mu$m and 102 mJy at 550~$\mu$m). 
\item A confusion-limited survey of about 100 square degrees.
To reach a 5$\sigma_{inst}$ noise limit of 14.1, 22.4 and 17.8 mJy,
one needs 192, 90 and 149 days at 250, 350 and 550~$\mu$m respectively.   
\item A very deep survey, down to the confusion limit to extract
as much information as possible about the underlying population.
For example, mapping 8 square degrees to 5$\sigma_{inst}$= 7.5 mJy
at 350 $\mu$m would take 64 days.
\end{enumerate}
We concentrate, in the following, on the 350 $\mu$m band (although
{\it SPIRE} is observing simultaneously the three bands).
The number of detected sources for each survey is given in Table \ref{Sens_SPIRE}
and the redshift distribution in Fig. \ref{z-350}.

\par\bigskip\noindent
$\bullet$ The very large-area survey\\
This survey is well suited to the {\it Planck}
sensitivity (Table \ref{Planck-tbl}). It would provide better positions
for the {\it Planck} point sources and the combination with the {\it Planck}
data would improve the spectral
and spatial characterisation of foregrounds. It will take about 18 days, which
is a small amount of time comparing to the enormous progress it would bring
for the component separation problem for all observations
in this wavelength range.
The very-large-area survey will detect mostly
nearby sources but hundreds of object could be detected at z$>$1.
It will also explore the cirrus component.
Moreover, thanks to its surface coverage, this survey will detect the rare and high-z objects
for which little is known today.

\par
\begin{figure*}
\begin{center}
\epsfxsize=10cm
\epsfbox{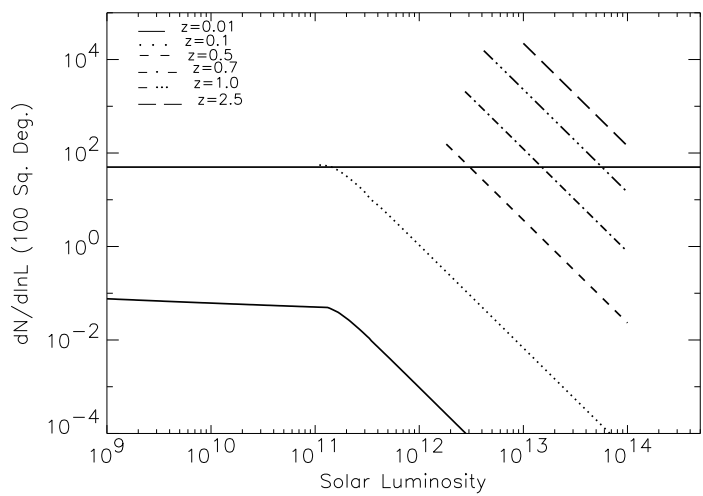}
\caption{\label{dn-350} Number of starburst galaxies that can be detected in different redshift
ranges (with a $\Delta$z/z=0.5) at 350 $\mu$m by the confusion limited survey of 100 square degrees
as a function of bolometric luminosity. The horizontal
line shows the 50 sources needed in a $\Delta$z/z=0.5 bin for $\sim$14$\%$ accuracy. The plots are limited to the fluxes above the
detection limit and to luminosities below 10$^{14}$ L$_{\odot}$.}
\end{center}
\end{figure*}

\bigskip\noindent
$\bullet$ The confusion-limited survey:\\
For the confusion-limited survey, there are mainly two peaks in the
redshift distribution: one at z$\sim$0 due to the cold
sources, the other between 0.8 and 3 being due to the starburst galaxies.
The number of starburst galaxies detected between 0.8 and 3 doesn't vary
much.
Fig. \ref{dn-350} predicts the number of starburst galaxies that the confusion-limited
survey will detect at 350 $\mu$m per log-interval of luminosities (this figure
does not include the cold population which contributes mostly locally).
The evolution of the 10$^{12}$-10$^{13}$ L$_{\odot}$ galaxies will be measured
from z$\sim$ 0.5 to 2.5. Galaxies with L$\sim$3 10$^{11}$ L$_{\odot}$
will be only accessible at z$\sim$0.1. 
The nature and number of the highest luminosity objects are important
to test cosmological theories, and thus populating the high redshift bins
with enough objects to ensure $<$14$\%$ Poisson noise (corresponding to 50 sources) 
is a critical driver
for the size of the sample, and accordingly, the area required for the survey.
Based on our model, we see from Fig. \ref{dn-350} that to meet
this $\sim$14$\%$ goal for galaxies in the range 10$^{12}$ - 10$^{13}$ L$_{\odot}$
at z$<$2.5, an area of 100 square degrees is the minimum required.

\par\bigskip\noindent
$\bullet$ The very-deep survey\\
The confusion-limited surveys will detect only $\sim$7$\%$ of the CIB
(1$\%$ for the vey-large area survey). In this case, studying the fluctuations
will bring informations on the underlying source population.
The redshift distribution of sources
making the bulk of the CIB and the fluctuations at 350 $\mu$m are very similar.
Thus, the fluctuations will be the unique
opportunity to get information, particularly on sources 
that dominate the CIB (sources with L$\sim$3 10$^{11}$ L$_{\odot}$), 
at redshifts where they cannot be detected individually.
To detect and study them, a field with 
high signal-to-noise ratio is needed. 
Moreover, the size of the field should be large enough
to try to detect the source clustering and not
only be limited to the Poissonian noise detection 
(which is about 4300 Jy$^2$/sr). Since the source clustering
is expected at the scales between 1 and 5 degrees (Knox et al. 2001),
a field of about 8 square degrees is the minimum required.
This field could be part of the confusion-limited survey
which will be certainly split in several smaller area surveys.

\subsubsection{The {\it PACS} surveys}
For the whole field of view of $\sim$1.75' $\times$ 3.5',
the current estimates of time needed to reach 5$\sigma_{inst}$=3 mJy
is 1 hour (Albrecht Poglitsch, private communication). 
At 75, 110 and 170~$\mu$m, the confusion limits are about
0.13, 0.89, and 7.08 mJy, with q$_{density}$ of 8.9, 8.7
and 7.1 respectively. To reach the confusion limit
for one field of view, i.e 5$\sigma_{inst}$=0.1, 0.9 and 7.1,
we need 567, 11 and 0.18 hours at 75, 110 and 170 $\mu$m respectively.
Since for {\it PACS} the confusion limits and the time to reach
those sensitivities are very different
at the three wavelengths, 
three kinds of surveys could be done that, schematically, 
will probe the CIB in the three bands:
\begin{enumerate}
\item A shallow survey of 20 square degrees down to the confusion
limit at 170 $\mu$m. This survey is dedicated to probe the CIB 
at 170 $\mu$m and study the correlations.
Such a survey will take 88 days. 
\item A deep survey down to the confusion limit at 110 $\mu$m.
A good compromise between the covered surface and the time needed
is a field of 25'$\times$25' that will take about 67 days. 
\item An ultra-deep survey down to the confusion limit at 75 $\mu$m
(or a little bit below the confusion limit). To map a 5'$\times$5'
field, 96 days are needed.
\end{enumerate}
These three surveys correspond to about the same amount of time
as the {\it SPIRE} surveys. Obviously, the {\it PACS} surveys have to be done within the
same areas as the {\it SPIRE} ones.\\

{\it PACS} is observing simulaneously the 170 and 110~$\mu$m channels
or the 170 and 75~$\mu$m channels. Ideally a combination
75/110 $\mu$m and 75/170 $\mu$m would have been preferable
since the 170~$\mu$m observation in the deep and ultra-deep
surveys will not bring new science compared to the
large area survey. On the contrary, observing the shortest wavelengths
in the shallow survey will detect the very luminous, hot and rare galaxies that
may be missed in the deep and ultra-deep surveys. \\

\begin{figure*}
\begin{center}
\epsfxsize=10cm
\epsfbox{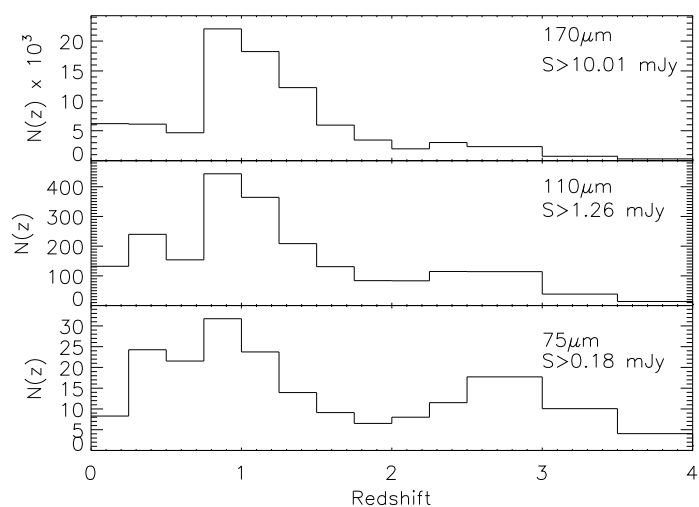}
\caption{\label{z-PACS} From bottom to top: redshift distribution for the ultra-deep, deep and shallow {\it PACS}
 surveys at 75, 110 and 170~$\mu$m respectively.}
\end{center}
\end{figure*}

The three surveys will detect thousand of sources at z$\sim$1 (Fig. \ref{z-PACS})
and will probe most of the CIB source population (they will resolve
about 49, 77 and 87$\%$ of the CIB at 75, 110 and 170 $\mu$m respectively, see 
Table \ref{Sens_PACS}). \\
At 170 $\mu$m, the shallow survey will give an unprecedent measurement
of the evolution of the 10$^{11}$-10$^{12}$ L$_{\odot}$ galaxies 
from z$\sim$ 0.25 to 1 and the evolution of the 10$^{12}$-10$^{13}$ L$_{\odot}$
galaxies from z$\sim$ 0.5 to 3 (with enough objects to ensure $<$10$\%$ Poisson noise).
Since half of the background is resolved into discrete sources at 170 $\mu$m, one
complementary approach is to reduce the surface of the shallow survey
to have a better signal-to-noise ratio and thus study the underlying
population. However, to study the correlation in the IR background,
a minimum of 8 Sq. Deg. is required. A survey at S$_{min}$ = 8 mJy, corresponding
to 5$\sigma_{inst}$=3.7 mJy, would take around 128 days for 8
Sq. Deg. 
Such a survey would give less statistics for the resolved sources
but would help in understanding the whole CIB population. 
With a 8 Square Deg. field, the measure of the high luminosity
source evolution would still be possible with a high
degree of accuracy.\\

In conlusion, {\it PACS} will resolve about 80$\%$ of the CIB around 100 $\mu$m
({\it SIRTF} will resolve at most around 55$\%$
of the CIB at 70 $\mu$m and 20$\%$ at 160 $\mu$m, Dole et al. 2002).
It will definitely resolve the question
of the population making the CIB near its emission's peak.
It will measure with unprecedented accuracy the history
of the IR-traced star formation up to z$\sim$1.5.
For the higher redshifts, the informations will
come mainly from the {\it SPIRE} surveys. Although {\it SPIRE}
will resolve less than 10$\%$ of the CIB in
the submm, it will bring an unprecedented information
on the evolution of galaxies up to z$\sim$3 
and also on the underlying population that
will be very hard to detect from the ground
due to the small area of the present surveys.


\subsection{The {\it Planck} all-sky survey}
For the {\it Planck} wavelengths, the confusion limits are given by the photometric
criterion{\footnote{The density criterion leads to $q_{density}$ from 1.5 to 0.8
from 350 to 2097~$\mu$m respectively.}}. With q$_{phot}$=5, they are about
447, 200, 79.4, 22.4 and 11.2 mJy at 350, 550, 850, 1380 and 2097 $\mu$m respectively 
(for detection only, q$_{phot}$=3 is better and leads to the following confusion limits:
251, 112, 44.7, 14.1 and 6.31 at 350, 550, 850, 1380 and 2097 $\mu$m respectively).
The confusion limit (with q$_{phot}$=5) is above the 5$\sigma$ instrumental
noise at 350 $\mu$m, comparable at 550 $\mu$m and then below at 
longer wavelengths. It can be compared to the values given in
the {\it `HFI proposal for the Planck mission (1998)'}, Table 2.1.
In this Table, the confusion limits have been computed
using $\Omega$ rather than $\int f^2(\theta, \phi) d\theta d\phi$ in Eq. \ref{eq_sigma}.
This leads to a systematic overestimates of $\sigma_{conf}$ by a factor 1.33.
Correcting for this factor, the ratio of those estimates to the present ones
increases from 350 to 2097~$\mu$m from 1.2 to 3. \\

To compute the number of sources expected
in the {\it Planck} survey, we use a cut in flux equal to 5$\sigma_{tot}$ such as:
\begin{equation}
\sigma_{tot} = \sqrt{ \sigma_{inst}^2 + \sigma_{conf}^2 + \sigma_{add}^2}
\end{equation}
where $\sigma_{conf}$ for {\it Planck} is given by  $\sigma_{conf}$ = S$_{lim}$/5.
$\sigma_{add}$ is an additonal noise due to unresolved sources
with fluxes between S$_{lim}$ and 5$\sigma_{inst}$:
\begin{equation}
\sigma_{add}^2 =  \int f^2(\theta, \phi) d\theta d\phi \int_{S_{lim}}^{5\sigma_{inst}} S^2 \frac{dN}{dS} dS
\end{equation}
This additional term is only present when the 5$\sigma_{inst}$ is greater
than the confusion limit (as for example for the {\it SPIRE} very-large survey).\\

Final sensitivities on point sources, together with
the number of detected sources for {\it Planck} are given
in Table \ref{Planck-tbl}. At all wavelengths, the sensitivity of the survey 
is in the euclidian part of the number counts. {\it Planck} will not be able to
constrain, with the resolved sources, the evolution of the submm
galaxies but it will give an absolute calibration
of the bright number counts, that will not be provided by any other planned instruments.
Moreover, by covering the whole sky, it will probably detect the most spectacular dusty object
of the observable Universe, as the hyperluminous or the strongly lensed
starburst or AGN galaxies, as well as extreme sources not included
in the model.\\
Note however that the sensitivities
have been computed with an instrumental noise derived from a mean
integration time. With the {\it Planck} scanning strategy, some high
latitudes regions (where the cirrus contamination is low)
will be surveyed more deeply, leading to an instrumental
noise about 3 times lower. 
In such high redundancy parts of the sky, {\it Planck} survey will be limited by the confusion noise
(except at 2097 $\mu$m where the instrumental and confusion noises
are on the same order). In those regions, {\it Planck} 
will produce unique maps of the CIB fluctuations.
CIB anisotropies are mainly contributed by moderate to
high redshift star-forming galaxies, whose clustering properties and
evolutionary histories are currently unknown. {\it Planck} observations
will thus complement the future far-IR and submm telescopes
from ground and space that will perform deep surveys over small
area. These surveys will resolve a substantial fraction of the CIB 
but will probably not investigate the clustering of the submm galaxies
since it requires surveys over much larger areas.

\subsection{Requirements for future experiments for very large deep surveys
in the IR/submm/mm domain}
Since the far-IR and submm sources (1) are often associated with mergers or interacting galaxies
and (2) their energy ouput is dominated by high
luminosity sources at high redshift (which might be related
to the optical where the output is dominated by high brightness sources at
high redshift (Lanzetta et al. 2002)), they have to be studied in detail 
in the long wavelength range and at high redshift (z$\geq$3) as a tool to understand the merging process
and the physics of the first non linear structures.\\

Two kinds of requirements have to be meet by the post {\it SIRTF}, {\it Herschel} and {\it Planck}
surveys: 
\begin{enumerate}
\item They have to find the interesting high redshift sources
which are early mergers made of building blocks not affected
much yet by star formation and evolution. The number of such sources have
to be large enough to do statistical studies.
\item Once these sources are found, the future experiments have to have enough sensitivity 
and angular resolution to study them in details.
\end{enumerate}

To quantify the first requirement, we compute the surface needed
to detect more than 100 sources with luminosities of 3 10$^{11}$ and 3 10$^{12}$ 
L$_{\odot}$ in each redshift range (Table \ref{sfce_pred}). At high redshift, we
need surveys of about hundred of square degrees to find
enough high luminous objects to do statistical studies.
If enough area is covered, we need moreover a high sensitivity,
for example we have to reach 0.21 mJy at 850 $\mu$m for L=3 10$^{11}$ L$_{\odot}$ galaxies
(dominating the LF at z$\sim$1)
and 1.9 mJy at 850 $\mu$m (or 0.9 mJy at 1300 $\mu$m) for L=3 10$^{12}$ L$_{\odot}$ galaxies
(dominating the LF at z$>$2).
The fluxes of typical 3 10$^{12}$ L$_{\odot}$ galaxies at high-z are just, for the single-antenna
telescopes, at the confusion limit. With the future wide-field imaging
instruments on these telescopes, for example {\it SCUBA-2} (Holland et al. 2001) and {\it BOLOCAM} (Glenn et al. 1998),
hundred of square degrees at the 1 mJy level at 1300 $\mu$m
could be mapped in a reasonable amount of time. However, with 10/15 arsec
beam, it will be very difficult to make optical identifications
and follow-up observations at other wavelengths.\\
To reach the sensitivity of the 3 10$^{11}$ L$_{\odot}$ galaxies at high-z,
we need to have a very good angular resolution not to be limited by the confusion.
Table \ref{resol_pred} gives the angular resolution, together with
the telescope diameter needed to reach 10, 30, 60 and 80 $\%$ of
the CIB at 350, 850 and 1300 $\mu$m. To resolve 80$\%$ of the background at 1300 $\mu$m,
we need a telescope diameter of about 173 m (!) and a diameter of about 113 m (!)
at 850 $\mu$m and 23 m at 350 $\mu$m. In conclusion, finding the objects 
that are making the bulk of the CIB at long
wavelengths will be a very challenging task! 
This leaves to an open question: how we find these objects?
In the mid-IR, the Next Generation Space Telescope (NGST) will be a great tool.
However, NGST observations will have two limitations: (1) the redshift up to which the 
dust component can be observed is limited to 4-5 and (2) the stellar component can be observed at higher z
but the experience of combined optical/near-IR and far-IR observations
shows how it is difficult to identify the galaxies which have
most of their output energy in the Far-IR from optical and
near-IR data alone. 
Therefore, the alternative today is to make the interferometres 
efficient enough to carry out large surveys.\\

For the second requirement, i.e study in detail the physics of the objects,
observations have to have enough sensitivity
to observe sub-components of the merging objects at high z 
(a 3 10$^{10}$ L$_{\odot}$ sub-component at z=5 has
a flux of about 0.014 mJy at 350 $\mu$m and of
about 0.028 mJy at 850 $\mu$m). Moreover, an angular resolution of about 0.2
to 1 arcsec, together with spectroscopic capabilities are needed.
For that, the submm and mm interferometres are the only tools, as shown
by the recent spectacular observations of high-redshift quasar with the IRAM
interferometre (Cox et al. 2002). Lower luminosity sources will require
the {\it ALMA} (http://www.eso.org/projects/alma/, http://www.mma.nrao.edu/)
or SPECS/SPIRIT (Leisawitz et al. 2001) interferometres.

\subsection{The case of {\it ALMA}}
Thanks to the negative K-correction, high redshift sources are
accessible from the ground at submm and mm wavelengths.
{\it ALMA}, a synthesis radio telescope (about 64 12-metre diameter telescopes) 
that will operate at submm and mm wavelengths 
will image the Universe with unprecedented sensitivity and angular resolution from
the high-altitude Llano de Chajnantor, in northern Chile. 
It will be one of the largest ground-based astronomy project of the next decade after VLT/VLTI, 
and, together with the NGST, one of the two major new facilities for world astronomy coming 
into operation by the end of the next decade. 
{\it ALMA}, with its angular resolution, great sensitivity and
spectroscopic capabilities will reveal in detail,
in the high-z galaxies, the astrophysical processes at work.
Moreover, {\it ALMA} will be free of limitation due to source
confusion and will therefore allow very
faint galaxies to be detected.
In this section, we discuss mostly {\it ALMA} abilities to
find large enough samples of high redshift sources to do statistical studies
and probe the CIB source population. Of course {\it ALMA} will be also
used to study their structure and physics.\\

At 1300 $\mu$m, to find enough $\sim$ 3 10$^{11}$ L$_{\odot}$
high-z sources, we need to cover a surface of at least 5 square 
degrees at a 5$\sigma$ level of about 0.1 mJy (Table \ref{sfce_pred}). 
Such a survey will resolve about 50$\%$ of the CIB. To resolve
$\sim$80$\%$ of the CIB, one needs to reach a 5$\sigma$
level of about 0.02 mJy (Table \ref{resol_pred}). Therefore two kinds of surveys
could be considered: a large-area ($\sim$ 5 Sq. deg.)
and an ultra deep ($\sim$ 10 arcmin$^2$) survey. 
For both surveys, the compact configuration has enough resolution
not to be limited by the confusion. At 1300 $\mu$m,
a 5$\sigma$ detection of 2.3 mJy is reached in 1 sec for a beam area
of 0.16 arcmin$^2$ ('{\it ALMA} proposal for phase 2' and Blain 2001).
This gives for the two types of surveys: 
\begin{itemize}
\item The large-area survey: 5 Sq. Deg., 5$\sigma_{1300}$ = 0.1 mJy (50$\%$ of the CIB) \\
A 1 square degree field requires 22 500 pointings, each with 529
seconds of observations, for a total of 138 days. For 5 Sq. Deg., 690 days are needed, i.e.
1.9 years. 
\item The ultra-deep survey: 10 arcmin$^2$, 5$\sigma_{1300}$ = 0.02 mJy (80$\%$ of the CIB) \\
A 10 arcmin$^2$ field requires 625 pointings, each with 13225 seconds
of integration, for a total of 96 days.
\end{itemize}

In conclusion, if we want to achieve the two goals: (i) 
detect enough early mergers made of building blocks not affected
much yet by star formation and evolution and (ii) probe most
of the CIB source population at large wavelengths, we will
have to do extragalactic surveys with {\it ALMA} using a substantial
fraction of the time to find the sources. 
Such large surveys including the whole {\it ALMA} collaboration
would be much more efficient in terms of scientific progress
than smaller area surveys conducted by individual smaller teams.\\


\section{Summary}
We have developped a phenomenological model that constrains in a simple way the IR
luminosity function evolution with the redshift, and fits all the existing source counts and
redshift distribution, CIB intensity and for the first time CIB
fluctuations observations from the mid-IR to the submm range.
The model has been used to give some predictions for future {\it Herschel} deep
survey observations and the all-sky {\it Planck} survey.
It comes out that the planned experiments ({\it SIRTF}, {\it Herschel}, {\it Planck}) 
will be mostly limited by the confusion.
To find out a large number of objects that dominate the LF
at high redshift (z$>$2), future experiments need both the angular
resolution and sensitivity. This can be achieved in the submm only
thanks to interferometres such as {\it ALMA}. However, mapping
large fractions of the sky with high signal-to-noise ratio will 
take a lot of time (for example, 1.9 years to map 5 square degrees
that resolve 50$\%$ of the CIB at 1.3~mm with ALMA).

\par\bigskip\noindent
\par\bigskip\noindent
{\bf APPENDIX A:}
\par\medskip\noindent
We provide, in an electronic form through a web page{\footnote{http://www.ias.fr/PPERSO/glagache/act/gal$_{-}$model.html}},
a distribution of the model's outputs and programs (to be used in IDL) containing:
\begin{itemize}
\item The array dN/(dlnL.dz) as a function of L and z, dS/dz as a function of L and z
and S$_{\nu}$ in Jy for each luminosity and redshift, from 10 to 2000 $\mu$m 
and the evolution of the LF
for both the normal and the starburst populations 
(for $\Omega_{\Lambda}$=0.7, $\Omega_{0}$=0.3 and h=0.65)
\item Some usefull programs that compute the integral counts,
detected sources, CIB and fluctuations redshift distribution,
and the level of the fluctuations from the previous data cubes. 
\end{itemize}
\par\bigskip\noindent

Acknowledgements: HD thanks the {\it MIPS} project (under
NASA Jet Propulsion Laboratory subcontract \# P435236)
for support during part of this work 
and the Programme National de Cosmologie and the
Institut d'Astrophysique Spatiale for some travel funding.

\bsp 

\newpage
\begin{table*}
\begin{center}
\caption{Predicted CIB intensity at 15 $\mu$m ({\it ISOCAM} filter),
60 and 100 $\mu$m (IRAS filters), 170 $\mu$m
({\it ISOPHOT} filter), 350 $\mu$m (filter such as $\Delta$$\lambda$ / $\lambda$= 1/3) and
850 $\mu$m ({\it SCUBA} filter) compared to measurements.}
\label{CIB-tbl}
\begin{tabular}{|c|c|c|l|l} \hline 
$\lambda$ & Predicted CIB & Predicted CIB & Measured CIB \\ 
 $\mu$m & MJy/sr & Wm$^{-2}$sr$^{-1}$ &  Wm$^{-2}$sr$^{-1}$ \\ \hline
15 & 1.25 10$^{-2}$ & 2.5 10$^{-9}$ & $>$ 2.4$\pm$0.5 10$^{-9}$ & (1) \\ \hline
60 & 0.12 & 6.1 10$^{-9}$ & - & \\ \hline
100 & 0.35 & 1.1 10$^{-8}$ & $\sim$1.5 10$^{-8}$ & (2)\\ \hline
170 & 0.76 & 1.3 10$^{-8}$ & 1.4$\pm$0.3 10$^{-8}$ & (3)\\ \hline
350 & 0.76 & 6.5 10$^{-9}$ & 5.63$^{+4.30}_{-2.80}$ 10$^{-9}$ & (4)\\ \hline
850 & 0.20 & 6.9 10$^{-10}$ & 5.04$^{+4.31}_{-2.61}$ 10$^{-10}$ & (4)\\ \hline
\end{tabular}\\
(1) Elbaz et al. (2002)\\
(2) Renault et al. (2001)\\
(3) From Kiss et al. (2001) and extrapolation of DIRBE measurements \\
(4) From Fixsen et al. (1998) and Lagache et al. (1999)\\
\end{center}
\end{table*}

\begin{table*}
\begin{center}
\caption{Comparison of the predicted CIB fluctuations (for $S<S_{max}$ in Jy$^2$/sr)
and the observations.}
\label{Fluc-tbl}
\begin{tabular}{|c|c|c|c|c} \hline 
$\lambda$ ($\mu$m) & S$_{max}$ (mJy) & Observations (Jy$^2$/sr) & References & Model (Jy$^2$/sr)\\ \hline
170 & 1000 & $\sim$25000 & Sorel et al., in prep & 23694 \\ \hline 
170 & 250 & 13000$\pm$3000 & Matsuhara et al. 2000 & 15644 \\ \hline 
170 & 100 & 7400 & Lagache \& Puget 2000 &  11629 \\ \hline
100 & 700$^\ast$ & 5800$\pm$1000 & Miville-Desch\^enes et al. 2002 & 10307  \\ \hline 
90 & 150 & 12000$\pm$2000 & Matsuhara et al. 2000 &  5290 \\ \hline 
60 & 1000 & 1600$\pm$300 & Miville-Desch\^enes et al. 2002 & 2507  \\ \hline 
\end{tabular}\\
$^\ast$ Bright sources in Miville-Desch\^enes et al.
(2002) are removed at 60 and 100~$\mu$m using the cut of 1 Jy at 60~$\mu$m.
Since the 60~$\mu$m sources are mostly starburst galaxies, we estimate
that 1 Jy at 60~$\mu$m is equivalent to 0.7 Jy at 100~$\mu$m.
\end{center}
\end{table*}

\begin{table*}
\begin{center}
\caption{{\it PACS} and {\it SPIRE} 1$\sigma$ confusion noise and its associated flux limit (S$_{lim}$)
for the photometric criterion (q$_{phot}$=5) and the source density criterion
(with the equivalent q$_{density}$).}
\label{conf_H}
\begin{tabular}{|l|r|c|c|} \hline
& & $\sigma$ & S$_{lim}$ \\ 
& & (mJy) & (mJy)  \\ \hline
{\it PACS} 75  $\mu$m & q$_{phot}$=5.0 & 2.26 10$^{-3}$ & 1.12 10$^{-2}$ \\
           & q$_{density}$= 8.9 & 1.42 10$^{-2}$ & 1.26 10$^{-1}$ \\ \hline
{\it PACS} 110 $\mu$m & q$_{phot}$=5.0 & 1.98 10$^{-2}$ & 1.0 10$^{-1}$ \\ 
           & q$_{density}$= 8.7 & 1.02 10$^{-1}$ & 8.91 10$^{-1}$ \\ \hline
{\it PACS} 170 $\mu$m & q$_{phot}$=5.0 & 3.97 10$^{-1}$ & 2.00 \\ 
           & q$_{density}$= 7.13 & 9.93 10$^{-1}$ & 7.08 \\ \hline
{\it SPIRE} 250 $\mu$m & q$_{phot}$=5.0 & 2.51 & 12.6 \\ 
           & q$_{density}$=5.2 & 2.70 & 14.1 \\ \hline
{\it SPIRE} 350 $\mu$m & q$_{phot}$=5.0 & 4.4 & 22.4 \\ 
           & q$_{density}$= 3.6 & 3.52 & 12.6 \\ \hline
{\it SPIRE} 550 $\mu$m & q$_{phot}$=5.0 & 3.69 & 17.8 \\ 
           & q$_{density}$=2.5 & 3.18 & 7.94 \\ \hline
\end{tabular}\\
\end{center}
\end{table*}

\begin{table*}
\begin{center}
\caption{Designed surveys that could be done with {\it SPIRE} (Numbers are from the 350~$\mu$m channel).}
\label{Sens_SPIRE}
\begin{tabular}{|c|c|c|c|c|c|c|} \hline 
Surface & $5\sigma_{inst}$  & $5\sigma_{conf}$ & $5\sigma_{tot}$ & Days & Number of & $\%$ resolved\\ 
(Sq. Deg.) & (mJy)  & (mJy) & (mJy) &  & sources & CIB \\ \hline
400 & 100  & 28.2$^1$ & 103.9 & 18 & 4768 & 1 \\ \hline
100 & 15.3 & 22.4 & 27.1 & 192 & 33451 & 6.7 \\ \hline
8 & 7.5 & 22.4 & 23.6 & 64 & 3533  & 7.8 \\ \hline
\end{tabular}\\
\end{center}
$^1$ Unresolved sources below
5$\sigma_{inst}$= 100 mJy induce a confusion noise of $\sigma_{conf}$=5.63 mJy.\\
\end{table*}

\begin{table*}
\begin{center}
\caption{Designed surveys that could be done with {\it PACS}.}
\label{Sens_PACS}
\begin{tabular}{|l|c|c|l|l|l|l|} \hline 
Surface & $\lambda$ & Days$^a$ &  $5\sigma_{inst}$ & S$_{min}$$^b$ & Number of & $\%$ resolved\\ 
        & ($\mu$m) &       &   (mJy)           &  (mjy)    & sources   &  CIB \\ \hline
20 Sq. Deg. & 170 & 88 & 7.08 & 10.01 & 87 322 & 48.7 \\ \hline
625 Sq. Arcmin & 110 & 67 & 0.89 & 1.26 & 1955 & 77 \\ \hline
25 Sq. Arcmin & 75 & 96 & 0.13 & 0.18 & 192 & 87
\end{tabular}\\
\end{center}
$^a$ Depending on the scanning/chopping/beam switching strategy, there may be some overhead 
of about 20$\%$\\
$^b$ $S_{min}$ = $\sqrt{(5\sigma_{inst})^2 + S_{lim}^2}$ = $\sqrt{2} \times S_{lim}$ 
\end{table*}

\begin{table*}
\begin{center}
\caption{{\it Planck} sensitivities (5$\sigma_{inst}$), confusion limit (S$_{lim}$= 5$\sigma_{conf}$),
confusion induced by sources between S$_{lim}$ and 5$\sigma_{inst}$ (5$\sigma_{add}$), and total 5$\sigma$ noise
(which is the quadratic sum of the contribution from the detector noise and the extragalactic confusion noise).
Also given are the cold, starburst and total galaxy densities.}
\label{Planck-tbl}
\begin{tabular}{|c|c|c|c|c|c|c|c|} \hline 
$\lambda$ & 5$\sigma_{inst}$ & 5$\sigma_{conf}$ & 5$\sigma_{add}$ & 5$\sigma_{tot}$ & N$_{cold}$(S$>5\sigma_{tot}$) & N$_{SB}$(S$>5\sigma_{tot}$) & N(S$>5\sigma_{tot}$) \\
($\mu$m) & (mJy)             & (mJy)           & (mJy) &       (mJy) & (/sr) & (/sr) & (/sr) \\ \hline
350 & 216.5 & 447 &  0 & 497 & 1342 & 40 & 1382  \\ \hline
550 & 219 & 200  & 7.9 & 297 & 187 & 15 & 202 \\ \hline
850 & 97 & 79.4 & 3.2 & 125 & 72 & 8 & 80 \\ \hline
1380 & 57.5 & 22.4 & 2.6 & 62 & 35 & 4 & 39  \\ \hline
2097 & 41.5 & 11.2 & 2.4 & 43 & 23 & 3 & 26  \\ \hline
\end{tabular}\\
\end{center}
\end{table*}

\begin{table*}
\begin{center}
\caption{Sky area in square degrees to be covered to detect more than 100 sources
in a redshift range $\Delta$z/z=0.3, for 3 10$^{11}$ L$_{\odot}$ (top) 
and 3 10$^{12}$ L$_{\odot}$ (bottom) starburst galaxies. }
\label{sfce_pred}
\begin{tabular}{c|l|l|l|l|l|} \hline 
3 10$^{11}$ L$_{\odot}$ & z&  S$_{350}$ (mJy) &  S$_{850}$ (mJy) & S$_{1300}$ (mJy) & Surface  \\ \hline
& 1 & 2.95 & 0.30 & 0.09 & 0.7 \\ \hline
& 3 & 0.78 & 0.25 & 0.08 & 1.8 \\ \hline
& 5 & 0.21 & 0.26 & 0.11 & 10 \\ \hline
& 7 & 0.065 & 0.21 & 0.12 & 60\\ \hline
 \end{tabular}\\
\begin{tabular}{c|l|l|l|l|l|} \hline 
3 10$^{12}$ L$_{\odot}$ & z&  S$_{350}$ (mJy) &  S$_{850}$ (mJy) & S$_{1300}$ (mJy) & Surface  \\ \hline
& 1 & 22.1 & 2.05 & 0.7 & 5.7 \\ \hline
& 3 & 7.51 & 1.82 & 0.6 & 6.1 \\ \hline
& 5 & 2.73 & 2.07 & 0.8 & 53 \\ \hline
& 7 & 1.01 & 1.89 & 0.9 & 398 \\ \hline
 \end{tabular}\\
\end{center}
\end{table*}

\begin{table*}
\begin{center}
\caption{Angular resolution and telescope diameter needed to have the confusion limit
at a flux level S$_{min}$ such that sources above S$_{min}$ contribute for
about 10, 30, 60 and 80 $\%$ of the CIB (top: 350 $\mu$m, middle: 850 $\mu$m, bottom: 1300 $\mu$m).}
\label{resol_pred}
\begin{tabular}{|l|l|l|l|l|} \hline 
$\%$ CIB & S$_{350}$ (mJy) & Log N (/sr) & $\theta^a$ (arcsec) & D$^b$ (metres) \\ \hline
80 &       0.9 & 8.01 & 3.2 & 23  \\ \hline
60 &       2.5 & 7.71 & 4.8 & 15.1  \\ \hline
30 &       8   & 7.09 & 10 & 7.2 \\ \hline
10 &       18  & 6.32 & 23.7 & 3.0 \\ \hline
\end{tabular}\\
\begin{tabular}{|l|l|l|l|l} \hline 
$\%$ CIB & S$_{850}$ (mJy) & Log N (/sr) & $\theta^a$ (arcsec) & D$^b$ (metres) \\ \hline
80 & 0.05 & 8.35 & 1.6 & 113  \\ \hline
60 & 0.2  & 8.03 & 3.0 & 59  \\ \hline
30 & 1    & 7.45 & 7.0 & 25  \\ \hline
10 & 2    & 6.77 & 15.3 & 12  \\ \hline
\end{tabular}\\
\begin{tabular}{|l|l|l|l|l} \hline 
$\%$ CIB & S$_{1300}$ (mJy) & Log N (/sr) & $\theta^a$ (arcsec) & D$^b$ (metres) \\ \hline
80         & 0.02           & 8.38        & 1.6                 & 173 \\ \hline
60         & 0.06           & 8.10        & 2.7                 & 101  \\ \hline
30         & 0.3            & 7.48        & 6.6                 & 40 \\ \hline
10         & 0.7            & 6.75        & 15.6                & 17 \\ \hline
\end{tabular}\\
$^a$ $\theta$ has been computed using N$\times \theta^2$ = 1/30 \\
$^b$ D = $\lambda$/ $\theta$ 
\end{center}
\end{table*}

\begin{figure*}
\label{LF_evol}
\begin{center}
\epsfxsize=18cm
\epsfbox{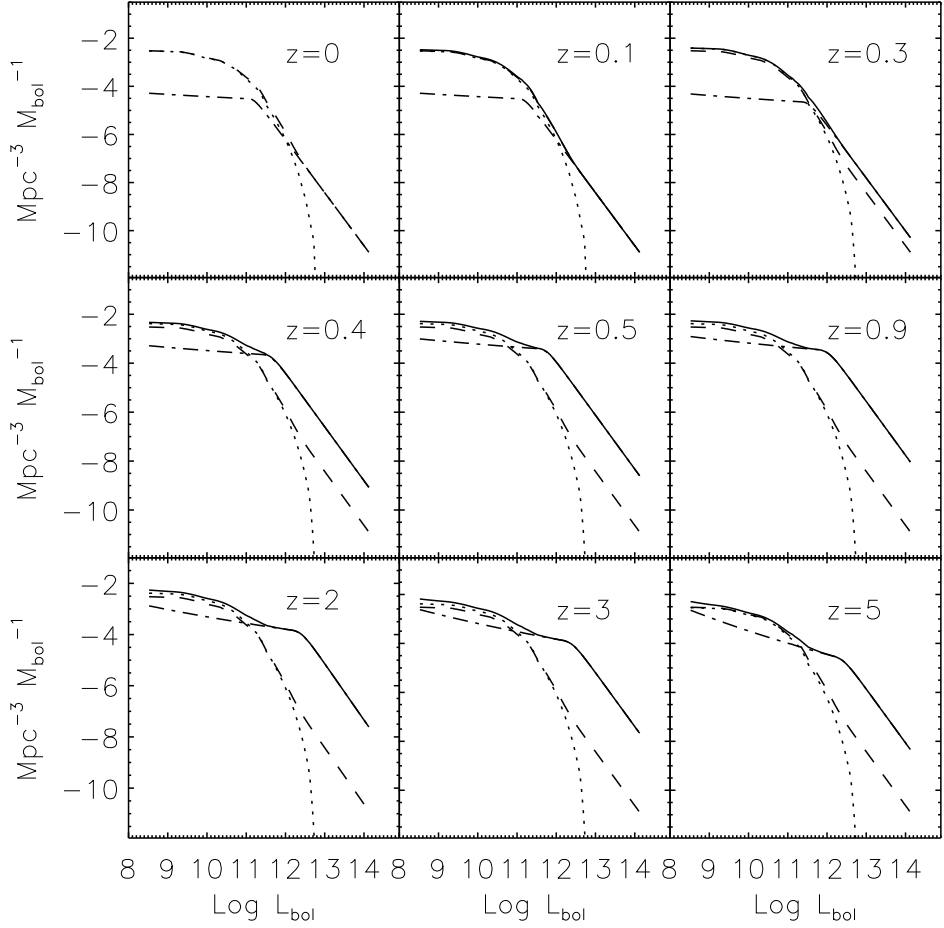}
\caption{\label{LF_evol} Co-moving evolution of the luminosity function. The dotted line is for the normal
galaxies and the dotted-dashed line, for the starburst galaxies. The continuous lines corresponds
to both starburst and normal galaxies and the dashed line is the LF at z=0 for comparison.}
\end{center}
\end{figure*}

\label{lastpage}

\end{document}